\documentclass[twocolumn]{aastex631}

\newcommand{\blue}[1]{\textcolor{blue}{#1}}

\usepackage{hyperref}
\usepackage[nolist]{acronym}
\usepackage{amsmath}
\usepackage{stmaryrd} 

\usepackage{CJKutf8}            

\usepackage{graphicx}
\usepackage{epstopdf}   
\usepackage{multirow}
\usepackage{booktabs}   
\usepackage{array}      
\usepackage{booktabs}   
\usepackage{makecell}   
\usepackage{diagbox}
\usepackage{threeparttable} 
\usepackage{appendix}
\usepackage[normalem]{ulem}  
\usepackage{comment}

\begin{document}

\title{Probing  Dark Matter Halos  of High-redshift  Quasars  via Wide-Field Clustering}

\author{Hao Meng \begin{CJK*}{UTF8}{gkai} (蒙皓) \end{CJK*}}
\affiliation{Department of Astronomy, Huazhong University of Science and Technology, Wuhan, Hubei 430074, China}

\author{Guangping Ye \begin{CJK*}{UTF8}{gkai} (叶广平) \end{CJK*}}
\affiliation{Department of Astronomy, Huazhong University of Science and Technology, Wuhan, Hubei 430074, China}

\author{Huanian Zhang \begin{CJK*}{UTF8}{gkai} (张华年) \end{CJK*}}
\affiliation{Department of Astronomy, Huazhong University of Science and Technology, Wuhan, Hubei 430074, China}
\affiliation{Hubei Fundamental Research Center for Physics, Wuhan, Hubei 430074, China}
\affiliation{Hubei Key Laboratory of Gravitation and Quantum Physics, Huazhong University of Science and Technology, Wuhan, Hubei 430074, China}
\email{huanian@hust.edu.cn}

\date{}

\begin{abstract}
High-redshift quasars are powerful tracers of both astrophysical processes and large-scale structure in the early Universe. Using a sample of 1,242 spectroscopically confirmed quasars and 827 highly reliable photometric quasar candidates selected with a machine-learning framework, all at $4.7 \leq z < 6.3$ (excluding $5.4 < z < 5.6$) and with a median luminosity of $M_{1450}\sim -25.5$, we investigate the large-scale environments of quasars near the end of cosmic reionization. We measure the projected auto-correlation function of the quasar population and derive bias parameters of $b=22.11^{+2.17}_{-2.19}$ and $28.55^{+5.76}_{-5.75}$ for the redshift intervals $4.7\leq z<5.4$ and $5.6\leq z<6.3$, respectively. These correspond to characteristic dark matter halo masses of $\log(M_{ h}/M_\odot)=12.66^{+0.10}_{-0.11}$ and $12.62^{+0.21}_{-0.21}$, respectively. We further estimate quasar duty cycles of $0.011^{+0.017}_{-0.007}$ and $0.012^{+0.104}_{-0.011}$ for the two redshift bins. These measurements are consistent with previous studies and the observed $f_{\rm duty}$--$M_{\rm halo}$ relation, indicating that only a small fraction of suitable dark matter halos host optically luminous quasars at any given time. The inferred low duty cycles suggest that a substantial fraction of SMBH growth may occur during obscured accretion phases. Our results provide new constraints on the connection between high-redshift quasars and their host dark matter halos, offering valuable insights into the co-evolution of SMBHs and large-scale structure in the early Universe.

\end{abstract}



\section{Introduction}\label{sec:intro}

Quasars are exceptionally luminous objects that can be observed even at very early Universe, with a few observed at $z > 7.5$ \citep{Banados2018, Yang2020, Wang2021}.  They are powered by gas accretion onto a supermassive black hole (SMBH) residing at the center of a galaxy, which is embedded within a dark matter halo (DMH). These SMBHs accrete gas from its surrounding environment, and those activities may trigger intermittent but violent outbursts, resulting in significant feedback effect that could regulate the growth of both the SMBH and the central galaxy \citep{Liu2025} and can ultimately affect the growth of the dark matter  halo \citep{QUO_SMBH_Salpeter_1964,QUO_SMBH_Lynden-Bell_1969,QUO_GLX_DMH_Kormendy_1995}.



The immediate environments of luminous high-redshift quasars are critically important for understanding the SMBH formation and its accretion. These environments may provide the exotic conditions necessary for forming massive black hole seeds \citep[e.g.,][]{Begelman2006, Regan2014, Inayoshi2020} and/or act as reservoirs for rapid, sustained gas accretion to fuel early growth \citep[e.g.,][]{DiMatteo2005}. To build a comprehensive model of early SMBH evolution $-$ encompassing their rapid growth, their connection to large-scale structure, and their co-evolution with host galaxies $-$ it is essential to observationally constrain the large-scale environments \citep[e.g.,][]{fan2006, Overzier2009, Drake2019} and the host DMHs of these quasars within the framework of cosmological structure formation \citep[e.g.,][]{Zhang2023}.

Statistically, the clustering analysis of quasars, commonly through the two-point auto correlation function \citep{LSS_Peebles_1980}, has proven to be
a powerful tool to connect the large-scale properties of quasars
with their hosting DMHs \citep{Shanks1994, Croom2001}. A key question about quasars and its corresponding large-scale environment is whether the most luminous quasars at high redshift, which trace the peaks of the density field, inhabit the same most massive DMH that will evolve into the cores of present-day rich clusters, or they represent transient, merger-driven phases in a moderate-mass systems. Therefore, studying the quasar's auto correlation function, inferring its DMH mass ($M_h$), as well as its evolution, offers a direct observational link to the life cycle of quasars and the assembly history of the massive DMH that serves as their foundational scaffolds. This approach bridges the gap between the small-scale physics of black hole accretion and the large-scale dynamics of structure formation in the Universe.

Moreover, to observationally constrain the growth scenarios of early SMBHs, two key quantities are essential: (1) the black hole masses during active phases, and (2) the fraction of their total growth that occurs during the UV-bright, luminous quasar phase we observe. The quasar clustering analysis provides a direct observational handle on this second quantity via the duty cycle, $f_{\rm duty}$. This is defined as the ratio of the observed quasar number density to the number density of their host DMHs: 
$f_{\rm duty} \equiv n_{\rm quasar}/n_{\rm halo} \sim t_{\rm Q}/t_{\rm H}$
where $t_{\rm Q}$ is the typical quasar lifetime and $t_{\rm H}$ is the Hubble time at the redshift of observation \citep{Efstathiou1988, Cole1989, HaimanHui2001, Martini2001}. Under the standard assumption that each halo hosts a single luminous quasar phase of duration $t_{\rm Q}$, $f_{\rm duty}$ represents the fraction of cosmic time that a typical halo shines as a quasar. A low duty cycle implies that most SMBH mass assembly must occur via radiatively inefficient or heavily obscured accretion, which contributes minimally to the observed UV luminosity. At high redshift ($z>5$), the quasar duty cycle is typically measured to be in the range of $\approx0.01\%$ to $1\%$, corresponding to quasar lifetimes of $t_{\rm Q} \approx 0.1$ to $10$~Myr. These estimates still suffer from large scatter due to uncertainties arising from cosmic variance, clustering analysis, attenuation, and the assumption of a single quasar per DMH.

At low-redshift epoch, the clustering analysis reveals that quasars are not randomly distributed but are preferentially located in regions of enhanced matter density \citep{DMHM_Croom_2005,DMHM_Shen_2007,Bias_Ross_2009,DMHM_Krumpe_2010,Shen2013,DMHM_Ikeda_2025,DMHM_Giner_Mascarell_2025}. They are found to be biased tracers of the underlying cosmic web, inhabiting the knots and filaments where dark matter collapses to form massive halos. The bias parameter, $b$, and the scale length, $r_0$, derived from the clustering analysis, are pivotal quantities, as they quantify the spatial concentration of quasars, and directly connect to the mass of the DMH where those quasars reside. Research on the bias parameter and DMH mass across cosmological times can provide a basic insight into the coevolution of quasars and the large-scale environment they reside \citep[e.g.,][]{Volonteri2012}. Therefore, the direct probe of the bias parameter and DMH mass at early Universe is critical for us to fully understand the formation and evolution of SMBH and its host galaxies/halos. 

Since the launch of the James Webb Space Telescope \citep[JWST,][]{JWST2023}, clustering analyses of quasars or active galactic nucleus (AGNs) and subsequent estimates of the DMH masses at high redshift ($z > 5$) have become increasingly common. For example, \citet{DMHM_Arita_2025} obtained $\log_{10}(M_h/h^{-1}M_{\odot})=11.53^{+0.15}_{-0.20}$, while \citet{DMHM_Lin_2026} derived $\log_{10}(M_h/M_{\odot})=11.04^{+0.34}_{-0.32}$ at $5.0<z<6.0$ and $11.21^{+0.35}_{-0.32}$ at $3.9<z<5.0$ using AGN–galaxy cross-correlation. \citet{Huang_2026_DMHM} obtain $\log_{10}(M_{h,\min}/M_{\odot}) = 12.27_{-0.26}^{+0.21}$, and \citet{Wang_2026_DMHM} estimate $\log_{10}(M_{h,\min}/M_{\odot}) = 12.1_{-0.4}^{+0.3}$ for luminous quasars with typical $M_{1450} \sim -26.4$ at $z\sim 6.6$, based on quasar–[O~{\small III}] cross-correlation using 25 quasar fields from the ASPIRE (A Spectroscopic Survey of Biased Halos in the Reionization Era, GO2078; PI Wang) program \citep{Wang2023, Yang2023}. For much more luminous quasars with  $M_{1450}$ between $-26.63$ and $-29.14$ at $z\sim6.25$, \citet{DMHM_Eilers_2024} infer a higher minimum halo mass of $\log_{10}(M_{h,\min}/M_{\odot}) = 12.43^{+0.13}_{-0.15}$ from quasar–galaxy cross-correlation. JWST observations now enable similar estimates for the highest-redshift sources. For a little red dot (LRD) and two quasars at $z = 7.3$, \citet{DMHM_Schindler_2025, JT2025} find $\log_{10}(M_{h,\min}/M_{\odot}) =12.0^{+0.8}_{-1.0}$ and $\log_{10}(M_{h,\min}/M_{\odot}) =11.64^{+0.56}_{-0.64}$, respectively. Using a large sample of spectroscopically confirmed LRDs, \citet{ZhangC2026} studied the cosmic evolution of the large-scale environments and their host DMHs of LRDs.

Those pioneering studies of high-redshift AGNs, quasars and LRDs have provided first glimpses into their host DMH, yet these results remain scarce, widely scattered, and highly uncertain. The lack of consensus stems from two key limitations: the comparably small sample sizes and extremely small survey areas which are susceptible to cosmic variance \citep[e.g.,][]{Driver2010, Robertson2010, Ucci2021}. Those limitations can significantly bias the estimates of the bias parameter, the clustering strength and the derived DMH properties. 

To address these challenges, we present a wide-field clustering analysis of a large sample of high-redshift quasars in order to mitigate the above limitations.  The high-redshift quasar sample includes both spectroscopically confirmed high-redshift
quasars and highly reliable photometric candidates that are robustly identified using a machine learning (ML) methodology \citep{Ye_2024ApJS} by combining $g, r, z$ photometry from the Legacy Surveys Data Release 9 \citep[LS DR9, ][]{Dey_2019, LS_DR9_Schlegel_2021} and $W1, W2$ photometry from the Wide-field Infrared Survey Explorer (WISE) all-sky survey \citep{WISE_Wright_2010, WISE_Mainzer_2011}.  The machine learning technique such as random forest and neural network is demonstrated to be quite effective in similar studies to search exotic objects using the same dataset \citep[e.g.,][]{ZhangDwarf2025}. The unprecedented combination of a large celestial area and a large data sample can effectively mitigate the  cosmic variance, providing more reliable insights into the cosmic environment of high-redshift quasars, the properties of their host DMHs, and their co-evolution history.

The paper is organized as follows. In Sec. \ref{sec:Data}, we present the data sample, including both spectroscopically confirmed high-redshift quasars and photometrically selected candidates, and the random sample used in this study. We also demonstrate the reliability of those photometric quasar candidates. In Sec. \ref{sec:Clustering Analysis}, we present the clustering analysis, obtain the projected auto correlation function, derive the DMH mass, the bias parameter, as well as the quasar duty cycle from the clustering analysis. In Sec. \ref{sec:dis}, we present the results using spectroscopically confirmed quasars only, the quasar luminosity $-$ halo mass relation, the potential cosmic evolution of the quasar DMH mass, and quasar duty cycles. We summarize and conclude in Sec. \ref{sec:Conclusions}. During this work, we adopt the cosmological model with
$H_0 = 67.66 ~\rm km ~s^{-1}~ Mpc^{-1}$, $\Omega_0 = 0.30966$ and $w_0 = -1$ \citep{Planck2018}. The dimensionless Hubble constant, $h$, is defined to be $h= H_0/100$.

\section{Data} \label{sec:Data}

While only thousands of quasars at $z > 4.5$ have been spectroscopically confirmed, tens of thousands of additional candidates at these redshifts have been identified photometrically. In this work, we therefore define our high-redshift quasar sample to include both the spectroscopically confirmed objects and the subset of photometric candidates deemed most reliable.

\subsection{Spectroscopically Confirmed High-redshift Quasars}

We compile a sample of spectroscopically confirmed high-redshift quasars in the redshift range $4.7 \leq z < 6.3$ from the literature \citep{Banados2016, Ross2020, Lyke2020, Matsuoka2022, Banados2023, known_quasar_fan_2023, known_quasar_yang_2023, Byrne2024, Belladitta2025, Matsuoka2025, MR2026, Yang2026}, including the comprehensive compilations of \citet{known_quasar_fan_2023} and \citet{known_quasar_yang_2023}, together with quasars identified in the Dark Energy Spectroscopic Instrument Data Release 1 \citep[DESI DR1;][]{DESIDR1_2025}. The resulting compilation contains 1,793 spectroscopically confirmed quasars at $4.7 \leq z < 6.3$. 

Because the spectroscopically confirmed high-redshift quasars are compiled from multiple surveys with heterogeneous selection criteria, we impose additional luminosity cuts to construct a more homogeneous sample and to minimize potential selection biases relative to the photometric quasar candidates introduced below. Specifically, we exclude quasars fainter than $M_{1450}=-24.5$, where $M_{1450}$ denotes the absolute magnitude at rest-frame 1450\,\AA\ and serves as a proxy for the ultraviolet (UV) luminosity. For quasars lacking published $M_{1450}$ measurements, we require a $z$-band magnitude brighter than 22, which approximately corresponds to this luminosity threshold at $z\sim5$ or $z\sim6$. After restricting the sample to the Legacy Surveys DR9 footprint and excluding the redshift interval $5.4 < z < 5.6$, where high-redshift quasars are notoriously difficult to identify photometrically because their broad-band colors overlap significantly with those of M-dwarf stars \citep{yang2017discovery}, the final sample comprises 1,242 quasars spanning the redshift range $4.7 \leq z < 6.3$, as summarized in Table \ref{tab:sample}.

\subsection{High-redshift Quasar Candidates}\label{subsec: high-zData}

The photometric high-redshift quasar candidates are drawn from \citet{Ye_2024ApJS}, who developed a detailed random forest (RF) framework for identifying high-redshift quasars. Here we briefly summarize the methodology. Using photometric measurements in the optical $g$, $r$, and $z$ bands from the DESI Legacy Imaging Surveys DR9, together with the near-infrared $W1$ and $W2$ bands from {\it unWISE} forced photometry, the authors constructed two parallel RF classifiers. The Legacy Surveys DR9 cover approximately $\sim14,000~\mathrm{deg}^2$ of the sky \citep{Dey_2019}, while the {\it unWISE} photometry is obtained by performing forced measurements on the {\it WISE} images using the optical source positions from the Legacy Surveys.

The two RF classifiers, referred to as the ``mag model'' and the ``flux model'', both achieve excellent classification performance. The ``mag model'' uses the $g$, $r$, $z$, $W1$, and $W2$ magnitudes as input features, whereas the ``flux model'' uses the corresponding flux measurements (see Section~3 of \citealt{Ye_2024ApJS} for details). Throughout this work, we adopt the ``flux model'' because it is more robust to missing data and low signal-to-noise measurements. This advantage is particularly important for high-redshift quasars, as strong absorption by neutral hydrogen blueward of Ly$\alpha$ often results in non-detections or even negative measured fluxes in the $g$- and/or $r$- bands. Indeed, more than half of known high-redshift quasars exhibit such measurements, making the flux model better suited for constructing a complete and reliable quasar candidate sample.

The flux-based classifier is trained on a spectroscopically confirmed sample comprising of 588 high-$z$ quasars, $\sim 1,000$ T-type and $\sim 4,500$ L-type brown dwarfs, as well as several million lower-redshift quasars and Galactic stars. It identifies a total of 568,188 candidates as well as the predicted probabilities of being a true high-redshift quasar from the classification model in the entire LS DR9 footprint. On a test set, the classification model achieves 95\% precision and 87\% recall. External validation using Multi Unit Spectroscopic Explorer \citep[MUSE,][]{MUSE2010} and DESI Early Data Release \citep[EDR,][]{DESI-EDR} spectra confirms its robustness; for instance, it recovers 9 of 10 new high-$z$ quasars in DESI EDR (90\% completeness on this unseen data). 

The photometric redshifts of those quasar candidates are estimated by a random forest regression model in the same work \citep{Ye_2024ApJS}.
The regression model is trained on $\sim$1,000 spectroscopically confirmed quasars with $4.5 < z < 6.5$, using the same photometric flux measurements as the classification model for the model inputs. 
On the test set, this model exhibits good performance (R-squared, $R^2$: 0.912; and mean squared error, MSE: 0.031). $R^2$ assesses how well the model explains variable changes (0 to 1, closer to 1 for better fit), while MSE directly reflects prediction accuracy (smaller values for higher accuracy). Hereafter, we divide the entire quasar sample into two redshift bins to better investigate their properties: a lower-redshift bin ($4.7 \leq z <5.4$, median $z \approx 5.0$) and a higher-redshift bin ($5.6 \leq z <6.3$, median $z \approx 6.0$). 

DESI is a state-of-the-art multi-object spectrograph instrument that deploys 5000 robotic fibers over a 3.2 degree diameter field of view \citep{Levi2013, DESI2016a, DESI2016b, DESIover}. DESI DR1 \citep{DESIDR1_2025} has already  collected spectra for tens of millions of astronomical objects, consisting of 13.1 million galaxies, 4 million Milky Way stars, and 1.6 million quasars, which deepens our understanding of the Universe and provides a great opportunity to verify the performance of the random forest classification model presented in \citet{Ye_2024ApJS}. We extract the spectra from DESI DR1 database, requiring 
$\texttt{zwarn} = 0$, $ \texttt{main\_primary} = \rm True$, and $\texttt{zcat\_primary} = \rm True$, which helps to select all unique, non-sky targets with no known redshift-fitting failures. 

From an initial photometric selection of high-redshift quasar candidates, we identified 3,157 objects with spectroscopic observations in DESI DR1 within 1$^{\prime\prime}$. According to the DESI classification, among these, 1,429 are compact galaxies, 1,190 are Milky Way stars, and 538 are quasars. However, only 383 of the 3,157 objects are true high-redshift quasars at $z>5$ after visual inspection, yielding a success rate far below the requirement for clustering analysis. Notably, the success rate $-$ defined as the fraction of confirmed high-redshift quasars among all DESI DR1 observations of our candidates $-$ increases significantly with the predicted probability. With these 3,157 objects with DESI spectroscopic observations on hand, we test various selection criteria to select a highly reliable subsample of high-redshift quasar candidates from the full candidate pool. 

\subsection{Removing Artifacts and Low-Quality Candidates}\label{subsec: Artifacts and Low-Quality}

First we find that there exist a certain fraction of artifacts or low photometric quality sources among the high-redshift quasar candidates after visual inspection. To address this, we apply a set of screening criteria based on Legacy Surveys parameters to remove such sources from the above selected candidate sample. These criteria are as follows:

\begin{enumerate}
\item \textbf{Continuous Observation.} A corresponding source detection in Data Release 10 (DR10) must exist within a 1.0 arcsec radius of the DR9 candidate. The southern sky region in DR10 spans a significantly wider observational time baseline compared to DR9. By cross-matching the coordinates of candidates between DR9 and DR10, we can effectively filter out transients and spurious artifacts. This is because genuine astrophysical sources are expected to persist across multiple epochs, whereas transient events or instrumental artifacts are unlikely to appear consistently at the same position in both datasets.

\item \textbf{Point-like source.} The source morphological type (\texttt{TYPE}) must be ``PSF" (point spread function) in both Legacy Surveys DR9 and DR10. Because DR10 incorporates additional photometric measurements and more observational epochs in certain regions, it provides more accurate morphological classifications. Enforcing this criterion allows us to remove extended sources or artifacts that were misclassified as ``PSF" in DR9 due to limited data quality.

\item \textbf{Bitmasks screening.} We exclude candidates flagged with any of the following \texttt{FITBITS} (2, 3, 4, 10, 11, 13) or \texttt{MASKBITS} (1, 5, 6, 7, 10, 12, 13).  
The \texttt{FITBITS} parameter indicates how well a source is fitted, while the \texttt{MASKBITS}\footnote{\url{https://www.legacysurvey.org/dr10/bitmasks/}} parameter identifies sources falling within the contaminated regions of bright foreground objects or specific structures. Filtering based on these bitmasks helps us strictly exclude sources with poor photometric reliability.

\item \textbf{Number of observations.} The number of $z$-band observations in DR9 must be greater than one (\texttt{nobs\_z} $> 1$). Via visual inspection, we found that the vast majority of artifacts among the candidates have only one or zero $z$-band observations (\texttt{nobs\_z} $\leq 1$). Therefore, we impose the \texttt{nobs\_z} $>1$ criterion. As a consequence, one spectroscopically confirmed high-redshift quasar is inadvertently excluded from the final sample.
\end{enumerate}

Applying these screening criteria reduces the sample size from 568,188 to 459,545 candidates, including 68,907 candidates at $4.7 \leq z < 5.4$ and 350,499 candidates at $5.6 \leq z < 6.3$.

\subsection{Mitigation of Targeting Bias}
\label{subsec: mitigation}

Then we find that using the DESI-observed subsample to directly infer the properties of the much larger unobserved candidate population introduces severe targeting biases, as DESI DR1 spectroscopy preferentially targets high-redshift quasar candidates that satisfy the traditional color-selection criteria of \citet[Yang23,][]{known_quasar_yang_2023}. To mitigate this effect, we combine the machine-learning selection with expanded color-selection criteria that incorporate additional near-infrared information. Specifically, we apply the Yang23 color cuts to the ML-selected parent sample; these cuts rely on Pan-STARRS DR1 \citep[PS1;][]{PS1dr12016} $i_{\rm PS1}$ and $y_{\rm PS1}$ photometry, which is not available in the Ye24 flux model. The Yang23 color-selection criteria adopted for the two redshift intervals are described below. Noting that our ML model separates quasars at $z < 5.0$ from the high-$z$ class ($5.0 < z < 6.5$), making it challenging to include the candidates at $4.7 < z < 5.0$ due to boundary effects. Therefore, only the photometric candidates within the redshift interval $5.0 \leq z < 5.4$ are included in the subsequent clustering analysis.

\textbf{1. Combined Selection Cuts for $5.0 \le z < 5.4$:}
\begin{itemize}
    \item \textbf{PS1 Detection \& LS Depth:} ${\rm S/N}(i_{\rm {PS1}}) > 5$ and $z_{\rm LS} < 21.4$.
    \item \textbf{Lyman Break Cut:} $0.8 < r_{\rm LS} - i_{\rm {PS1}} < 3.0$.
    \item \textbf{Extended Continuum Color Cuts:} 
    \begin{align*}
        i_{\rm {PS1}} - z_{\rm LS} &< 0.6 \times (r_{\rm LS} - i_{\rm {PS1}}) - 0.2; \\
        -0.5 &< i_{\rm {PS1}} - z_{\rm LS} < 0.75
    \end{align*}
    \item \textbf{Additional $y$-band Constraint:} If valid PS1 $y_{\rm {PS1}}$ photometry exists, the object must satisfy either (${\rm S/N}(y_{\rm {PS1}}) < 3$) or ($y_{\rm {PS1}} - W1_{\rm LS} > -0.3$ and $z_{\rm LS} - y_{\rm {PS1}} < 0.5$).
\end{itemize}

\textbf{2. Combined Selection Cuts for $5.7 \le z < 6.3$:}
\begin{itemize}
    \item \textbf{PS1 $i_{\rm {PS1}}$ Dropout:} Either ${\rm S/N}(i_{\rm {PS1}}) < 3$ or $i_{\rm {PS1}} - z_{\rm LS} > 2.0$.
    \item \textbf{Additional $y$-band Constraints:} If valid $y_{\rm {PS1}}$ photometry exists, the object must satisfy two logical conditions: (a) ${\rm S/N}(y_{\rm {PS1}}) < 3$ \textbf{OR} $y_{\rm {PS1}} - W1_{\rm LS} > -0.3$; \textbf{AND} (b) Either ${\rm S/N}(z_{\rm {PS1}}) < 3$, \textbf{OR} ${\rm S/N}(y_{\rm {PS1}}) < 3$, \textbf{OR} the object must concurrently satisfy the PS1 dropout rule (${\rm S/N}(i_{\rm {PS1}}) < 3$ or $i_{\rm {PS1}} - z_{\rm {PS1}} > 2.0$) \textbf{AND} a flat continuum constraint ($z_{\rm {PS1}} - y_{\rm {PS1}} < 1.0$).
\end{itemize}

\begin{figure}[!ht]
\includegraphics[width=.98\linewidth]{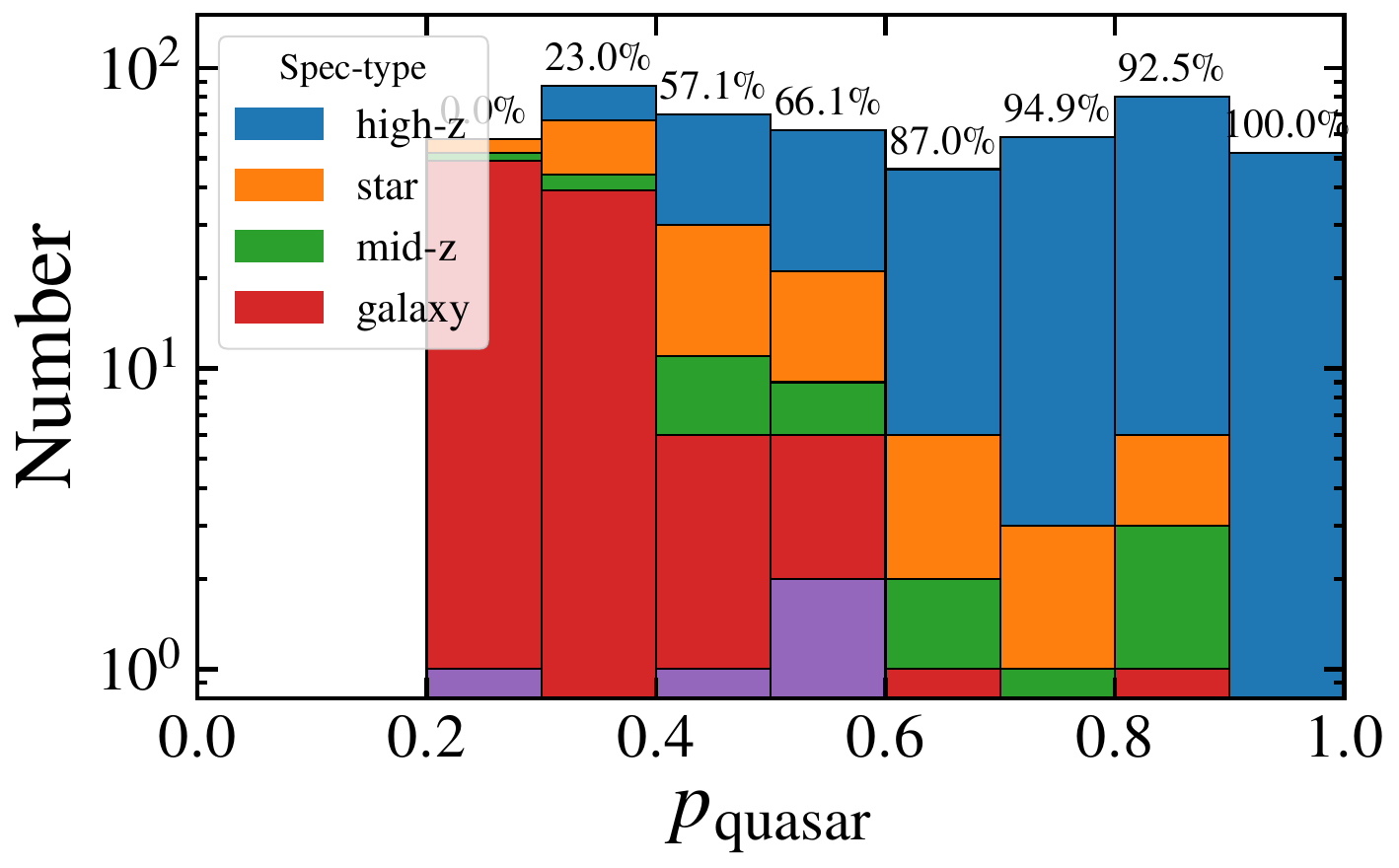}
\includegraphics[width=.98\linewidth]{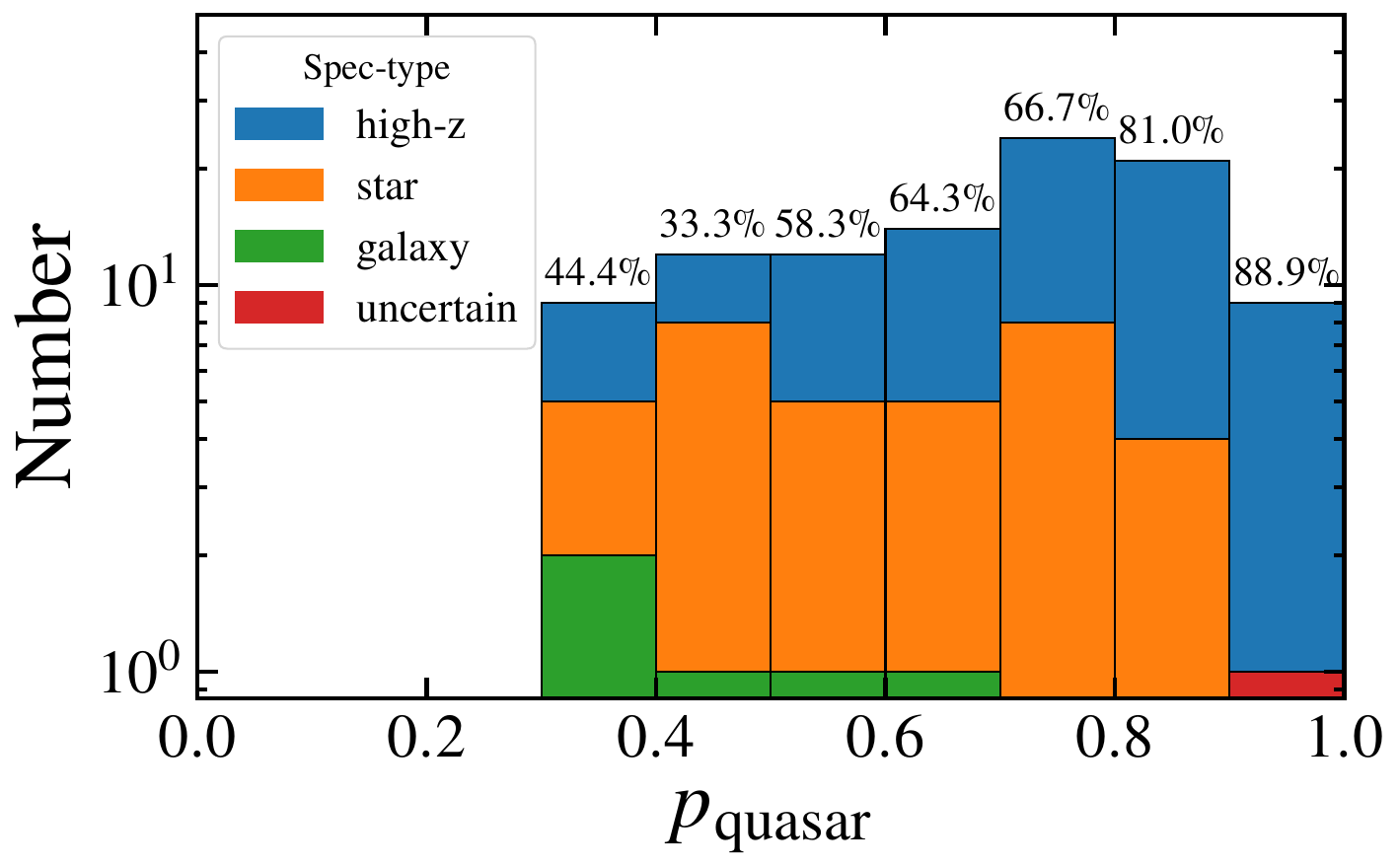}

\caption{ The classification performance of the candidates with DESI spectroscopic observations as a function of the predicted probability of the ``high-$z$'' class for the redshift interval of $4.7 \leq z <5.4$ (top, $z$-band magnitude $<21.4$) and $5.6 \leq z <6.3$ (bottom, $z$-band magnitude $<21.0$). The fraction shown above each bar is the proportion of these ``high-$z$'' objects among all observed candidates in the corresponding bin.}
\label{fig:performance}
\end{figure}

After applying these color-selection criteria, we obtain 8,041 and 13,149 candidates in the lower- and higher-redshift bins, respectively, and substantially mitigate the targeting bias. However, these samples are still not ideal for clustering analyses because their expected success rates remain relatively low based on the spectroscopic follow-up statistics reported by \citet{known_quasar_yang_2023}. We therefore impose additional selection criteria to construct a higher-purity quasar candidate sample. For the redshift interval $4.7 \leq z < 5.4$, we find that requiring $p_{\rm quasar}>0.7$ and a $z$-band magnitude brighter than 21.4 yields a success rate of approximately 96\% (182/190), as shown in the top panel of Figure~\ref{fig:performance}. This panel illustrates the classification performance of DESI-observed candidates as a function of the predicted probability of belonging to the high-redshift quasar class. For the redshift interval $5.6 \leq z < 6.3$, the criteria $p_{\rm quasar}>0.8$ and $z$-band magnitude brighter than 21.0 achieve a success rate of 83\% (25/30), as shown in the bottom panel of Figure~\ref{fig:performance}. This panel presents the corresponding classification performance for DESI-observed candidates in the higher-redshift bin. The $z$-band magnitude cut ensures that these photometric quasars satisfy the $M_{1450}$ threshold.


After removing sources with existing spectroscopic observations, the final photometric sample comprises 757 candidates at $4.7 \leq z < 5.4$ and 70 candidates at $5.6 \leq z < 6.3$, as summarized in Table~\ref{tab:sample}. We note that photometric candidates are not included in the redshift interval $4.7 < z < 5.0$ because of boundary effects in the classification model. Consequently, only spectroscopically confirmed quasars are retained in this redshift range.  The absolute magnitudes at rest-frame 1450~\AA, $M_{1450}$, are estimated using either spectroscopic or photometric redshifts (Ye, Zhang et al., submitted). Here we briefly describe the procedure used to derive $M_{1450}$. To obtain consistent luminosity estimates across the full redshift range ($4.7 \leq z < 6.3$), we follow the widely used photometric method of \citet{Banados2016}. Specifically, the apparent magnitude at rest-frame 1450~\AA, $m_{1450}$, is estimated by extrapolating either the LS $z$-band or PS1 $y_{\rm PS1}$-band photometry, assuming a quasar continuum described by a power law, $f_\nu = C\nu^{\alpha_\nu}$, with a spectral index of $\alpha_\nu=-0.3$ \citep{Selsing2016}. Using an alternative spectral index of $\alpha_\nu=-0.5$ \citep{Vanden2001} produces negligible differences in the inferred $m_{1450}$ values, as demonstrated by \citet{Banados2016}. The corresponding absolute magnitudes, $M_{1450}$, are then calculated from $m_{1450}$ using the spectroscopic or photometric redshift of each source. 

\begin{figure}[!ht]
    \centering
    \includegraphics[width=\linewidth]{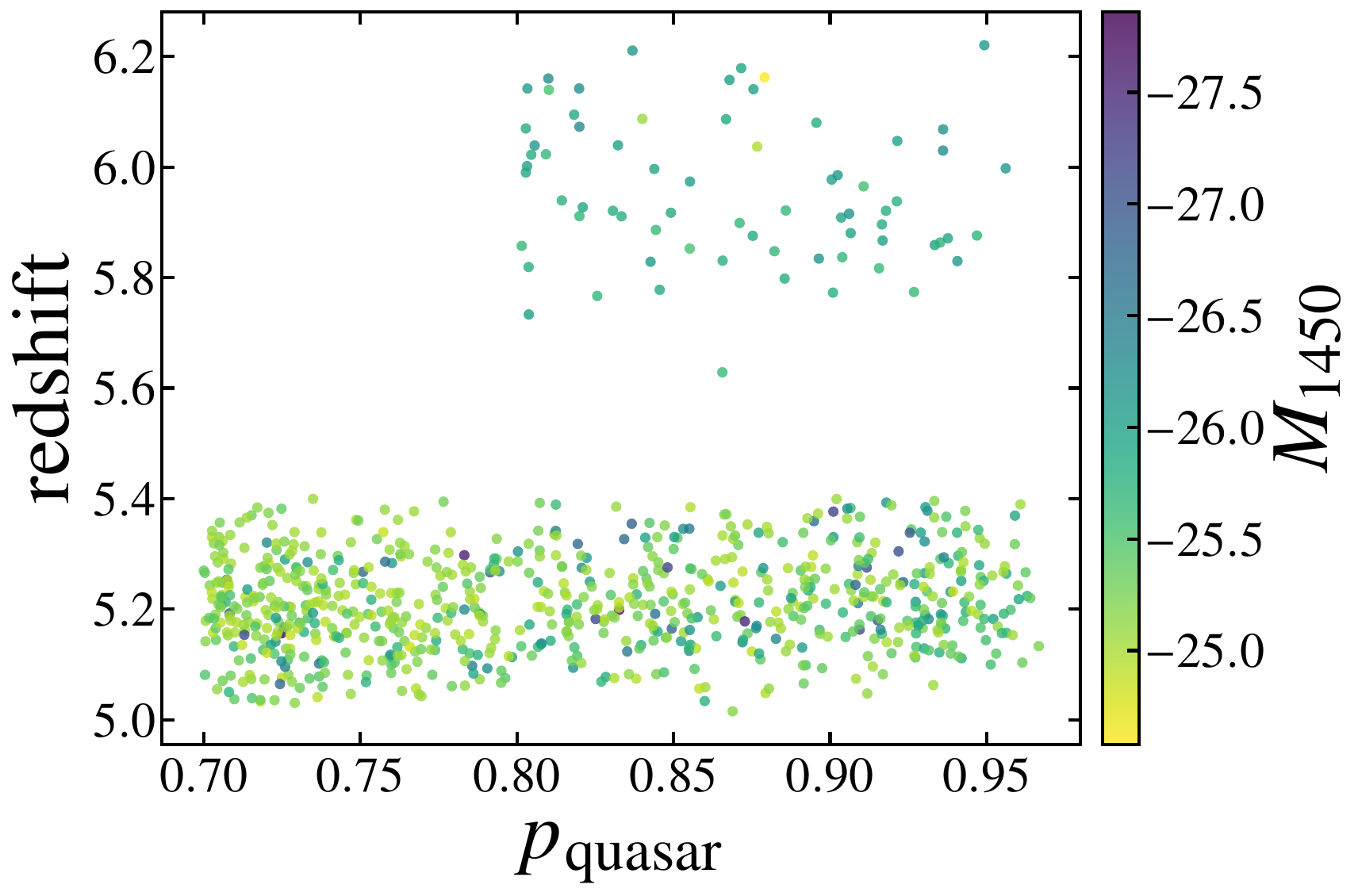}
    \caption{The distribution of the redshift and predicted probability of the final photometric candidate sample. The color coding denotes the absolute magnitude, $M_{1450}$.}
    \label{fig:pquasar_z}
\end{figure}

Figure~\ref{fig:pquasar_z} shows the distribution of the photometric quasar candidates in the $p_{\rm quasar}-z$ plane, with points color-coded by their absolute magnitude $M_{1450}$. The broad distribution of $M_{1450}$ across the full range of $p_{\rm quasar}$ values indicates that the classifier selects candidates over a wide luminosity range rather than being dominated by the most luminous objects. Furthermore, unlike the full candidate pool, whose number counts decline rapidly with increasing $p_{\rm quasar}$ (Figure~9 of \citealt{Ye_2024ApJS}), the final candidate sample exhibits a nearly uniform distribution in $p_{\rm quasar}$. This difference indicates that our selection preferentially retains sources with consistently high quasar probabilities, lending further confidence to the reliability of the resulting candidate sample.

\begin{figure*}[!ht]
    \centering
    \includegraphics[width=0.96\linewidth]{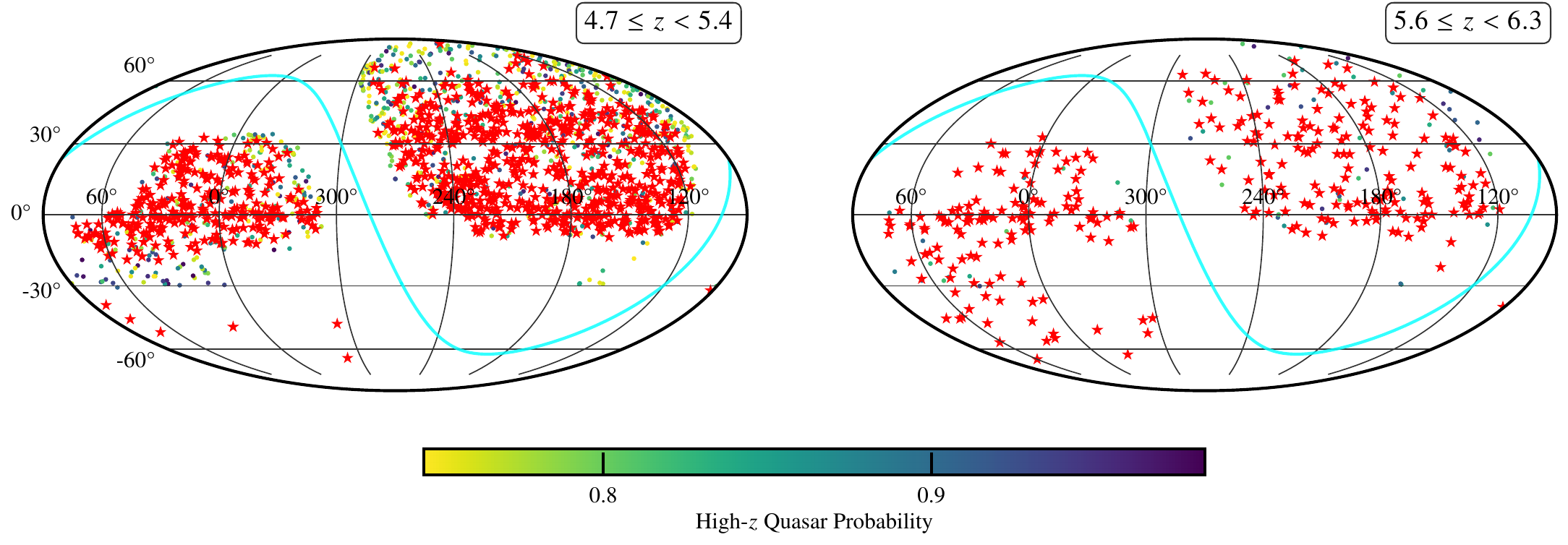}
    \caption{ The sky distribution of quasars at $4.7 \leq z <5.4$ (left) and $5.6 \leq z <6.3$ (right). The color coding denotes the predicted probability of being a true high-$z$ quasar obtained in \citet{Ye_2024ApJS} for the candidates. Red stars represent the spectroscopically confirmed quasars within the same redshift range.
    Cyan solid line represents the Galactic plane. 
    }
    \label{fig:sky}
\end{figure*}

The resulting quasar samples have median UV luminosities of $M_{1450}=-25.4$ and $-25.9$ for the lower- and higher-redshift bins, respectively. The sky distributions of both the spectroscopically confirmed quasars and the photometric quasar candidates are shown in Figure~\ref{fig:sky}.

\begin{table}[!ht]
\setlength{\tabcolsep}{6pt}
\caption{The number of spectroscopically confirmed quasars and highly reliable candidates for the two redshift bins.}
\centering
\begin{tabular}{c|cc}
\hline
Sample & \multicolumn{2}{c}{Number} \\ 
                & $4.7 \leq z <5.4$   & $5.6 \leq z <6.3$  \\ \hline
Confirmed       & $941$               & $301$       \\
Candidates      & $757$        & $70$       \\
Total           & $1,698$             & $371$        \\ \hline
\end{tabular}

\label{tab:sample}
\end{table}

We construct the random catalog using the Legacy Surveys Data Release 9 (DR9) random catalogs\footnote{\url{https://www.legacysurvey.org/dr9/files/}}. These random points were generated across the Legacy Surveys footprint on a brick-by-brick basis, where each brick covers approximately $0.0623 \ \mathrm{deg}^2$, with a mean surface density of $\sim2500\ \mathrm{deg}^{-2}$. For each random position, survey metadata were extracted from the corresponding {\it coadd} images, enabling the random catalog to accurately trace the survey geometry and observational footprint (see the Legacy Surveys documentation for details). The random points are distributed according to a homogeneous Poisson process \citep{random_Myers_2023} and are designed to reproduce the angular selection function of the Legacy Surveys. To construct a three-dimensional random catalog, redshifts are assigned to the random points such that the resulting redshift distribution matches that of the data sample. Ideally, the random catalog should reproduce all selection effects present in the data sample. However, our high-redshift quasar compilation is assembled from multiple surveys with heterogeneous target-selection criteria and survey depths, making it difficult to impose a fully consistent selection function on the random catalog. As discussed above, to mitigate potential biases arising from these differences, we restrict the clustering analysis to luminosity-limited quasar samples, requiring $M_{1450}<-24.5$ for all spectroscopically confirmed quasars. These luminosity thresholds are at least 1 mag brighter than the practical sensitivity limits of the Legacy Surveys and help ensure that the analyzed quasars are drawn from a regime where the survey completeness is relatively high and reasonably uniform across the survey footprint. The resulting random catalog is therefore expected to provide an adequate representation of the survey geometry and selection function for the luminosity-limited quasar samples used in this work.

\section{Clustering Analysis}\label{sec:Clustering Analysis}

We first estimate the  auto correlation function of high-redshift quasars in Sec. \ref{subsec:wp}, by measuring the projected correlation function $\omega_p(r_p)$ of quasars in the two redshift bins. Meanwhile we estimate their uncertainties and derive the corresponding real space correlation function $\xi(r)$, which is assumed to be a power-law function characterized by the correlation length ($r_0$) and power-law index ($\gamma$).  Then we derive the bias parameter in Sec. \ref{subsec:Bias}. The DMH mass and the duty cycle will be estimated based on the bias parameter in Sec. \ref{subsec:DMH Mass} and Sec. \ref{subsec:duty cycle}, respectively.

\subsection{Projected Auto Correlation Function}\label{subsec:wp}
The quasars' three-dimensional auto correlation function quantifies the clustering effects of quasar, which is easily affected by the redshift-space distortions (RSD) in observation.
To mitigate RSD, we compute the two-dimensional projected correlation function $\xi(r_p, \pi)$, where $r_p$ and $\pi$ represent the decomposition of the three-dimensional separation between two objects ($s$) into components perpendicular and parallel to the line of sight, respectively (i.e., $s=\sqrt{r_p^2+\pi^2}$). Various estimators exist for the projected correlation function, and we adopt the estimator from \citet{2DACF}:
\begin{equation}
\xi(r_p, \pi) = \frac{DD(r_p, \pi)-2DR(r_p, \pi)+RR(r_p, \pi)}{RR(r_p, \pi)}
\label{eqa:ACF}
\end{equation}
where $DD(r_p, \pi),~DR(r_p, \pi),~RR(r_p, \pi)$ represent the counts of data-data pair, data-random pair and random-random pair, respectively.

We adopt \texttt{Corrfunc} \citep{Corrfunc2020}\footnote{\url{https://github.com/manodeep/Corrfunc}}, a Python package that provides routines for performing pair counting in clustering analysis. The projected correlation function $\omega_p(r_p)$ is derived by
integrating $\xi(r_p, \pi)$ within $\pi$ direction, as follows:
\begin{equation}
\omega_p(r_p)=\int_{-\pi_{\rm max}}^{\pi_{\rm{max}}}\xi(r_p, \pi)\rm{d}\pi = 2 \int_{0}^{\pi_{\rm{max}}}\xi(r_p, \pi)\rm{d}\pi
\end{equation}
where $\pi_{\rm{max}}$ is the optimum limit above which the
clustering signal is almost negligible. In this research, we adopt  $\pi_{\rm{max}}=80\ h^{-1}\rm{Mpc}$,  consistent with the values in \citet{DMHM_Arita_2023}. 

There are different methods to estimate the statistical uncertainties of the auto correlation function measurement, either internally using bootstrap or jackknife resampling or externally using mock catalogs based on cosmological simulations. Here we adopt the jackknife resampling method \citep{Error_ACF} to estimate the uncertainty of the auto correlation function measurements $\omega_p(r_p)$. We divide the entire quasar sample as well as the random sample into $N=20$ supsamples of roughly equal celestial area, and eliminate each  subsample in turn to measure the correlation function for the remaining samples.  The covariance error matrix is then estimated as follows:
\begin{equation}
\begin{aligned}
        C_{ij} =&\frac{N-1}{N}\sum^{N}_{k=1}(\omega_{p,k}(r_{p,i})-\bar{\omega}_p(r_{p,i}))
        \\&\times(\omega_{p,k}(r_{p,j})-\bar{\omega}_p(r_{p,j})),
    \label{eqa:covmatrix}
    \end{aligned}
\end{equation}

where $\omega_{p,k}(r_{p,i})$ and $\bar{\omega}_p(r_{p,i})$ denote the projected correlation function measured after excluding the $k$-th subsample and the mean value across all such subsamples, respectively, for the $i$-th $r_p$ bin. The covariance matrix is generally dominated by its diagonal elements. The uncertainty in $\omega_p(r_{p,i})$ is therefore taken as the square root of the corresponding diagonal element, i.e., $\sigma_i = \sqrt{C_{ii}}$.


The projected correlation function is derived from the integration of the real–space correlation
function $\xi(r)$ \citep{RSCorr} as follows:

\begin{equation}
    \omega_p(r_p) = 2 \int^{\infty}_{r_p}\frac{r\xi(r)}{\sqrt{r^2-r_p^2}}dr,
\end{equation}
where the real-space correlation function is normally assumed to follow a power-law function with a correlation length ($r_0$) and a power-law index ($\gamma$):
\begin{equation}
    \xi(r)=\left(\frac{r}{r_0}\right)^{-\gamma},
    \label{eqa:RSCorr}
\end{equation}
Therefore, the fitted correlation function $\omega_{p,\rm{fit}}(r_p)$ is represented analytically with a beta function $B$ as follows: 
\begin{equation}
    \frac{\omega_{p,\rm{fit}}(r_p)}{r_p} = B\left(\frac{\gamma-1}{2},\frac{1}{2}\right)\left(\frac{r_p}{r_0}\right)^{-\gamma}.
\end{equation}

\begin{figure*}[!ht]
    \centering
    \includegraphics[width=\linewidth]{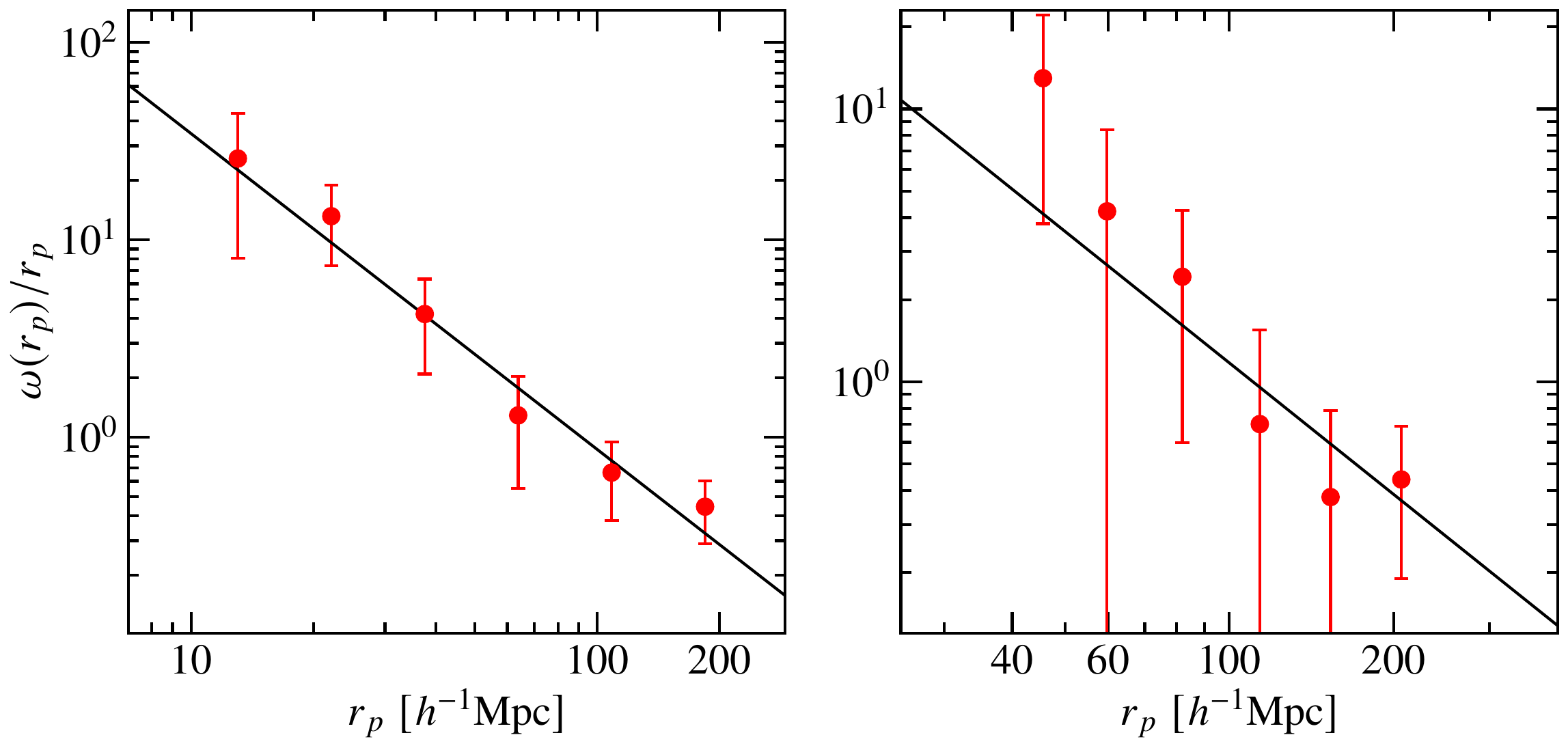}
    \caption{Projected auto correlation function for quasars at $4.7 \leq z <5.4$ (left) and $5.6 \leq z <6.3$ (right). The red dots represent the measurements of the projected auto correlation and the black solid line stands for the best-fit model.} 
        \label{fig:ProjCorr}
\end{figure*}

Due to the limited sample size, obtaining robust constraints with a fully free power-law slope is challenging. We initially fit the measured projected correlation functions by allowing $\gamma$ to vary as a free parameter; however, this resulted in large uncertainties in both the clustering parameters and the inferred DMH properties. We therefore fix $\gamma=1.60$ for both redshift bins. This value is consistent with the $1\sigma$ confidence intervals of the $\gamma$ values obtained from multiple free-$\gamma$ fitting trials. Moreover, a similarly shallow slope of the projected correlation function on large scales ($r_p \gtrsim 10~\mathrm{Mpc}$) has been observed in previous luminous quasar clustering studies at $z>3.5$ (e.g., Figure~7 of \citealt{DMHM_Shen_2007}). Such behavior is expected because the large-scale clustering signal is dominated by the two-halo term, where the correlation function departs from a simple power-law form and typically exhibits a flatter slope than on small scales. The measured projected correlation functions, $\omega_p(r_p)$, together with the best-fitting models, are presented in Figure~\ref{fig:ProjCorr} for quasars at $4.7 \leq z < 5.4$ and $5.6 \leq z < 6.3$, yielding best-fit correlation lengths of $r_0 = 35.49^{+4.29}_{-4.59}~h^{-1}\mathrm{Mpc}$ for the lower-redshift sample and $r_0 = 42.85^{+9.63}_{-11.32}~h^{-1}\mathrm{Mpc}$ for the higher-redshift sample. These measurements are consistent with $-$ though slightly higher than $-$ previous clustering analyses of high-redshift quasars \citep{DMHM_Arita_2023, Huang_2026_DMHM, Wang_2026_DMHM}. 

For comparison, adopting the canonical slope of $\gamma=2.0$ yields significantly poorer fits to the measured projected correlation functions. At $z\sim5$, the model deviates from several data points by more than $1\sigma$ and produces a correlation length approximately 30\% larger than our fiducial estimate. At $z\sim6$, the inferred correlation length is roughly 50\% larger and accompanied by substantially larger uncertainties. These results indicate that a steeper correlation-function slope is not favored by the current large-scale clustering measurements. Nevertheless, the limited sample size prevents a robust determination of $\gamma$. Future wide-area spectroscopic surveys with much larger high-redshift quasar samples will enable more precise clustering measurements, particularly on small scales ($r_p \lesssim 10\,h^{-1}\mathrm{Mpc}$), allowing direct constraints on $\gamma$, improved halo occupation distribution modeling, and tighter measurements of halo masses, quasar duty cycles, and quasar lifetimes, as discussed below.  


\subsection{The Bias Parameter}\label{subsec:Bias}

Quasars, as one of the most luminous objects in the Universe, are assumed to reside in the densest part of dark matter distribution and trace the distribution of underlying dark matter \citep{Bias_mass_function_Sheth}, with a certain bias. The bias parameter is defined to be the ratio of clustering strength between quasars and underlying dark matter at $r = 8 \ h^{-1}$ Mpc;
\begin{equation}
    b=\sqrt{\frac{\xi(8,z)}{\xi_{\rm{DM}}(8,z)}}
    \label{eqa:observed_bias}
\end{equation}
where $\xi(8,z)$ denotes the quasars' clustering strength and is approximated by the power-law function defined by Eq. \eqref{eqa:RSCorr}. $\xi_{\rm{DM}}(8,z)$ is the clustering strength of underlying dark matter and is usually estimated using \texttt{halomod} \citep{halomod2013, halomod2021} package\footnote{\url{https://github.com/halomod/halomod}}, assuming the  bias model developed by \citet{Tinker_2010}, the transfer function from \texttt{CAMB}\footnote{\url{https://github.com/cmbant/CAMB}} \citep{CAMB2011},  and the growth model proposed by \citet{growth_model}.

Based on the measured projected correlation function and the best-fit model, the bias parameters are estimated to be $22.12^{+2.19}_{-2.21}$
 for quasars at $4.7 \leq z <5.4$ and $28.70^{+5.56}_{-5.74}$ for quasars at $5.6 \leq z <6.3$. The latter is consistent with the values reported by \citet{DMHM_Arita_2023}. The uncertainty of $b$ is estimated by bootstrap uncertainty estimation with replacement.  We first generate a large amount of random samples of $(r_0, \gamma)$ assuming they follow Gaussian distributions with the mean and standard deviation estimated above, and then we repeat the bootstrap resampling 10,000 times and calculate $b$ as in Eq. \eqref{eqa:observed_bias}. From the distribution of $b$ measurements, we determine the mean value and we quote to use the 16 and 84 percentiles as the uncertainty range. 

\subsection{DMH Mass of  Quasars}\label{subsec:DMH Mass}

We derive the typical DMH mass from the bias parameters of the quasar's auto correlation function. Under the standard assumption that quasars are biased tracers of the underlying dark matter distribution, their effective bias $b(M_h, z)$ can be modeled via the halo occupation distribution (HOD) framework. Following \citet{Tinker_2010}, the large-scale bias for halos of mass $M_h$ at redshift $z$ is given by: 
\begin{equation}
    b(\nu)= 1-A\frac{\nu^a}{\nu^a+\delta^a_c}+B\nu^b+C\nu^c,
    \label{eqa:fittingBias}
\end{equation}
where $\nu$ is the peak height of the linear density field and defined as $\nu = \delta_c/\sigma(R)$. $\delta_c$
is the critical density for the collapse of DMHs, adopted with $1.686$ according to \citet{Tinker_2010}. $\sigma(R)$ is the linear matter variance on the Lagrangian scale of the halo, i.e., $R=(3M/4\pi\bar \rho_m)^{1/3}$, defined as

\begin{equation}
    \sigma^{2}(R)=\frac{1}{2 \pi^{2}} \int_{0}^{k_{\max }} P(k, z) \hat{W}^{2}(k, R) k^{2} \rm{d} k,
    \label{eqa:sigma}
\end{equation}
where $P(k, z)$ is the matter power spectrum at redshift $z$ generated by \texttt{CAMB} with the adopted cosmological parameters, and $R$ represents the radius
of the DMH $R_{\rm{halo}}$. $\hat{W}(k,R)$ is the Fourier transform of the top-hat window function of radius $R$. $k_{\max}$ is usually set to be $10\ h \ \rm{Mpc}^{-1}$ in order for the above integration to converge. $\hat{W}(k,R)$ is defined as follows:
\begin{equation}
    \hat{W}(k,R) = \frac{3}{(kR)^3}[\sin(kR)-kR\cos(kR)].
    \label{eqa:hatW}
\end{equation}


Solving Eq. \eqref{eqa:fittingBias} with the adopted parameters in Table 2 of \citet{Tinker_2010}, we can obtain $R_{\rm{halo}}$, from which we can derive the typical mass of the DMH as follows:
\begin{equation}
    M_{h,\rm typ} = \frac{4}{3}\pi\bar\rho_mR_{\rm{halo}}^3,
\end{equation}
where $\bar\rho_m$ is the mean density of the Universe and is calculated to be $2.78\times
10^{11}\Omega_mh^2M_{\odot}/\rm{Mpc}^3$. 

Based on the bias parameters derived from the projected auto-correlation function, we infer characteristic DMH masses of $\log(M_{h,\rm typ}/M_{\odot})=12.66^{+0.10}_{-0.11}$ for quasars at $4.7 \leq z < 5.4$ and $\log(M_{h,\rm typ}/M_{\odot})=12.62^{+0.21}_{-0.21}$ for quasars at $5.6 \leq z < 6.3$. The uncertainties in the halo mass estimates are derived using a bootstrap resampling approach and propagated from the corresponding uncertainties in the bias measurements. The inferred characteristic halo masses are consistent between the two redshift bins despite the substantial difference in cosmic time. This similarity is not unexpected given the comparable UV luminosity distributions of the two quasar samples and slightly higher median luminosity for quasars at the higher redshift bin. Furthermore, the large sky coverage and substantial comoving volume probed by our survey reduce the impact of cosmic variance, supporting the robustness of the inferred halo mass estimates.

\subsection{Duty Cycle and Lifetime of Quasars}\label{subsec:duty cycle}

In the standard model of SMBH growth \citep{Salpeter1964}, SMBHs are assumed to accumulate mass exponentially.  This scenario assumes that SMBH growth proceeds continuously, resulting in a simple exponential growth curve as follows:
\begin{equation}
    M_{\rm BH} = M_{\rm seed}\exp(t/t_S),
\end{equation}
where $M_{\rm seed}$ is the black hole seed mass; $t_S$ is the Salpeter time, characterizing the growth rate of black holes, and can be evaluated as follows:
\begin{equation}
    t_S=450\times\frac{\epsilon}{1-\epsilon}\frac{L_{\rm Edd}}{L_{\rm bol}} ~\mathrm{Myr}
\label{eqa:Salpeter Time}
\end{equation}
where $\epsilon$ is the radiative efficiency; $L_{\rm Edd}$ is the Eddington luminosity; $L_{\rm bol}$ is the bolometric luminosity.

Consequently, these massive SMBHs will appear as luminous quasars for a certain fraction of their cosmic history. However, this scenario ignores the intermittent growth process of black holes, as discussed in studies such as \citet{DMHM_Shen_2007, DMHM_Arita_2023, DMHM_Schindler_2025}. Following these works, we adopt the duty cycle $f_{\rm duty} \equiv t_{\rm Q} / t_H(z)$  for quasars, which represents the ratio of the quasar lifetime to the Hubble time $t_H(z)$,  in order to gain deeper insight into SMBH growth.


It is assumed that a DMH with a minimum mass threshold, $M_{h,\min}$, can host a quasar. The duty cycle is then equivalent to the ratio of the number of observed quasars with a minimum luminosity ($M_{1450,\min}$, hereafter $M_{\min}$) to the total number of host DMHs above $M_{h,\min}$, as follows:

\begin{equation}
    f_{\rm{duty}}= \frac{n_{\rm Q}}{n_{\rm halo}} = \frac{\int_{M_{ \text{min}}}^{\infty}\Phi(M)dM} {\int^{\infty}_{M_{h,\min}}n(M_h)dM_h}
    \label{eqa:fduty}
\end{equation}

We adopt $M_{\min} = -24.5$ for both redshift bins, as discussed above. The minimum DMH mass $M_{h,\min}$ is estimated from the measured bias parameter and will be discussed in detail in the following. Here $n(M_h)$ represents the DMH mass function at a certain redshift $z$. $\Phi(M)$ is the quasar luminosity function (QLF) at the same redshift $z$, which is well approximated by a broken double power-law \citep[i.e.,][]{Boyle2000}:
\begin{equation}
\Phi(M,z) = \frac{\Phi^*(z)}
{10^{0.4(\alpha+1)(M-M^{*})}
+ 10^{0.4(\beta+1)(M-M^*)}}
\label{eqa:QLF}
\end{equation}

It is defined by the normalization $\Phi^*(z)$, the break magnitude $M^*$, and the two power-law slopes $\alpha$ and $\beta$. The normalization factor $\Phi^*(z)$  takes the form of the following:

\begin{equation}
\Phi^*(z)= \Phi^*(z=6)\times 10^{ k(z-6)}
\end{equation}
where the parameter $k$ describes the exponential evolution of the quasar density with redshift. Typical values of $k$ are $-0.47$ \citep{k=0.47_Willott2010} or $-0.7$ \citep{k=0.7_Jiang2016}. The parameters of the QLF are adopted from \citet{Yang2016} for $4.7 \leq z < 5.4$ and from \citet{JT2023} for $5.7 \leq z < 6.2$, as summarized in Table~\ref{tab:QLFparam}.

\begin{table}[!ht]
\setlength{\tabcolsep}{1pt} 
\caption{The adopted parameters of the QLF. }
\centering
\begin{tabular}{@{}cccccc@{}}
\toprule
$z$ & $\log\Phi^*(z=6)$ & $M^*$  & $\alpha$ & $\beta$ & $k$ \\ \midrule
5.0 & $-8.82^{+0.15}_{-0.15}$ & $-26.98^{+0.23}_{-0.23}$ & $-2.03$ & $-3.58^{+0.24}_{-0.24}$ & $-0.47$\\
6.0 & $-8.75^{+0.47}_{-0.41}$ & $-26.38^{+0.79}_{-0.60}$ & $-1.70^{+0.29}_{-0.19}$ & $-3.84^{+0.63}_{-1.21}$ & $-0.70$\\ 
\bottomrule
\end{tabular}

\label{tab:QLFparam}
\end{table}

Here we adopt the DMH mass function proposed by \citet{Mass_Function_Tinker_2008}, as follows:

\begin{equation}
\frac{dn}{dM_h}=f(\sigma)\frac{\bar\rho_m}{M_h}\frac{d\ln\sigma^{-1}}{dM_h},
\label{eqa:dn_dM}
\end{equation}
\begin{equation}
    f(\sigma) = A\left[\left(\frac{\sigma}{b}\right)^{-a}+1\right]e^{-c/\sigma^2}.
    \label{eqa:f(sigma)}
\end{equation}

where $\sigma$ is the same as in Eq. \eqref{eqa:sigma}. The remaining parameters are adopted as $A=0.186$, $a=1.47$, $b=2.57$, and $c=1.19$, following Table 2 of \citet{Mass_Function_Tinker_2008}. It should be noted that this mass function (Eq. \eqref{eqa:dn_dM} and \eqref{eqa:f(sigma)}) was implemented using the Python package \texttt{COLOSSUS}\footnote{\url{https://pypi.org/project/colossus/}} by \citet{COLOSSUS2018}. The minimum halo mass is then estimated from the measured bias parameter as follows:
\begin{equation}
b_{\mathrm{eff}}=\frac{\int_{M_{h,\min}}^{\infty} b(M_h, z) n(M_h) d M_h}{\int_{M_{h,\min}}^{\infty} n(M_h) d M_h},
\end{equation}
where $b(M_h, z)$  is the bias parameter of a given DMH at a certain redshift $z$ as in Eq. \eqref{eqa:fittingBias} and $b_{\mathrm{eff}}$ is interpreted as the mean bias for DMHs whose masses exceed $M_{h,\min}$. 

We obtain \(\log(M_{h,\min}/M_{\odot}) = 12.38^{+0.11}_{-0.11}\) and \(f_{\rm duty} = 0.011^{+0.017}_{-0.007}\) for quasars at \(4.7 \leq z < 5.4\), and \(\log(M_{h,\min}/M_{\odot}) = 12.36^{+0.23}_{-0.22}\) and \(f_{\rm duty} = 0.012^{+0.104}_{-0.011}\) for quasars at \(5.6 \leq z < 6.3\), with the latter being consistent with previous estimates within uncertainties \citep{DMHM_Arita_2023, Huang_2026_DMHM, Wang_2026_DMHM}. According to the definition of duty cycle, \(f_{\rm duty} = t_{\rm Q}/t_H(z)\), we derive corresponding quasar lifetimes of \(t_{\rm Q} = 12^{+19}_{-8} ~\mathrm{Myr}\) and \(t_{\rm Q} = 11^{+99}_{-10}~\mathrm{Myr}\) for the two redshift bins, respectively, with all values summarized in Table~\ref{tab:compare}.

Standard scenarios assume that the SMBHs grow continuously from the seed with mass of $\sim 10M_{\odot}$ at $z = 35$ to $\sim10^{10}M_{\odot}$ at $z=6$ \citep{Inayoshi2020} with $\epsilon=0.1$. However, because the growth is intermittent with a duty cycle $f_{\rm duty}$, then to achieve the same final mass growth from the seed, the Salpeter time must be reduced by a factor of $f_{\rm duty}$ compared to the continuous case. Considering $\epsilon$ in the Eq. \eqref{eqa:Salpeter Time}, this reduction in active accretion time forces a much lower radiative efficiency: $\epsilon =0.12\%$ for $4.7 \leq z <5.4$ and $\epsilon= 0.13\%$ for $5.6 \leq z <6.3$, consistent with that of \citet{Huang_2026_DMHM}. Such low efficiencies might imply that these quasars are highly obscured at observed wavelengths \citep{Trebitsch2019}. 

\section{Discussion}\label{sec:dis}

\begin{table*}[!ht]
\centering
\caption{Clustering results from the combined sample including photometric candidates (All) and from the spectroscopically confirmed quasars (Spec), at $z\sim5.0$ and $z\sim6.0$, respectively. }
\setlength{\tabcolsep}{3pt}
\providecommand{\blue}[1]{\textcolor{blue}{#1}}
\begin{tabular}{l c c c c}
\toprule
 & \multicolumn{2}{c}{$z\sim5.0$} & \multicolumn{2}{c}{$z\sim6.0$} \\
\cmidrule(lr){2-5}
& All & Spec & All & Spec \\
\midrule
$r_0$ ($h^{-1}$Mpc) & 
$35.49^{+4.29}_{-4.59}$ & 
$28.67^{+7.14}_{-8.06}$ & 
$42.85^{+9.63}_{-11.32}$ & 
$46.87^{+13.86}_{-16.19}$ \\
$\gamma$ & 
$1.60$ & 
$1.60$ & 
$1.60$ & 
$1.60$ \\
$b$ & 
$22.12^{+2.19}_{-2.21}$ & 
$17.88^{+3.81}_{-3.85}$ & 
$28.70^{+5.56}_{-5.74}$ & 
$30.72^{+7.98}_{-8.33}$ \\
$\log(M_{\rm typ}/M_\odot)$ & 
$12.66^{+0.11}_{-0.11}$ & 
$12.47^{+0.25}_{-0.24}$ & 
$12.62^{+0.21}_{-0.21}$ & 
$12.65^{+0.30}_{-0.27}$ \\
$\log(M_{\rm min}/M_\odot)$ & 
$12.38^{+0.11}_{-0.11}$ & 
$12.17^{+0.27}_{-0.26}$ & 
$12.36^{+0.23}_{-0.22}$ & 
$12.40^{+0.32}_{-0.29}$ \\
$f_{\rm duty}$ & 
$0.011^{+0.017}_{-0.007}$ & 
$0.0023^{+0.0105}_{-0.0019}$ & 
$0.012^{+0.104}_{-0.011}$ & 
$0.014^{+0.218}_{-0.013}$ \\
$t_{\rm Q}$ (Myr) & 
$12^{+19}_{-8} $ & 
$3^{+13}_{-2} $ & 
$11^{+99}_{-10}$ & 
$13^{+207}_{-12}$ \\
\bottomrule
\end{tabular}
\label{tab:compare}
\end{table*}

Using the spectroscopically confirmed quasar sample alone, consisting of 941 quasars at $4.7 \leq z < 5.4$ and 301 quasars at $5.6 \leq z < 6.3$, we independently construct the projected auto-correlation functions and derive the corresponding clustering properties following the same methodology described in Section~\ref{sec:Clustering Analysis}. The resulting correlation functions are fully consistent with those obtained from the combined sample, which includes both spectroscopically confirmed quasars and high-confidence photometric candidates (Figure~\ref{fig:ProjCorr}). The best-fitting clustering parameters and the corresponding halo properties are summarized in Table~\ref{tab:compare}.

For the lower-redshift bin ($z\sim5$), the spectroscopic-only sample yields $r_0=28.67^{+7.14}_{-8.06}~h^{-1}\,\mathrm{Mpc}$, $b=17.88^{+3.81}_{-3.85}$, and $\log(M_{\rm typ}/M_\odot)=12.47^{+0.25}_{-0.24}$, all consistent within the uncertainties with the corresponding values derived from the combined sample. For the higher-redshift bin ($z\sim6$), where only 70 photometric candidates are added to the 301 spectroscopically confirmed quasars, the agreement is even closer. The spectroscopic-only sample yields nearly identical values of $r_0=46.87^{+13.86}_{-16.19}~h^{-1}\,\mathrm{Mpc}$, $b=30.72^{+7.98}_{-8.33}$, and $\log(M_{\rm typ}/M_\odot)=12.65^{+0.30}_{-0.27}$. The inferred minimum halo masses are likewise in excellent agreement, indicating that the inclusion of high-confidence photometric candidates does not introduce any measurable systematic bias into the halo-mass estimates.

The duty-cycle and quasar-lifetime estimates exhibit larger uncertainties owing to their stronger dependence on the halo mass function and the limited sample size. Nevertheless, they remain statistically consistent with the values derived from the combined sample.  Overall, the close agreement between the spectroscopic-only and combined analyses demonstrates that the inclusion of the high-confidence photometric candidates improves the statistical precision of the clustering measurements while preserving the inferred physical properties of the quasar population, thereby supporting the reliability of our candidate-selection procedure. This consistency also indicates that photometric-redshift uncertainties of those candidates do not introduce a significant systematic bias into our clustering results.

\subsection{DMH Mass Over Cosmic Time}

\begin{figure*}[!ht]
    \centering    \includegraphics[width=0.75\linewidth]{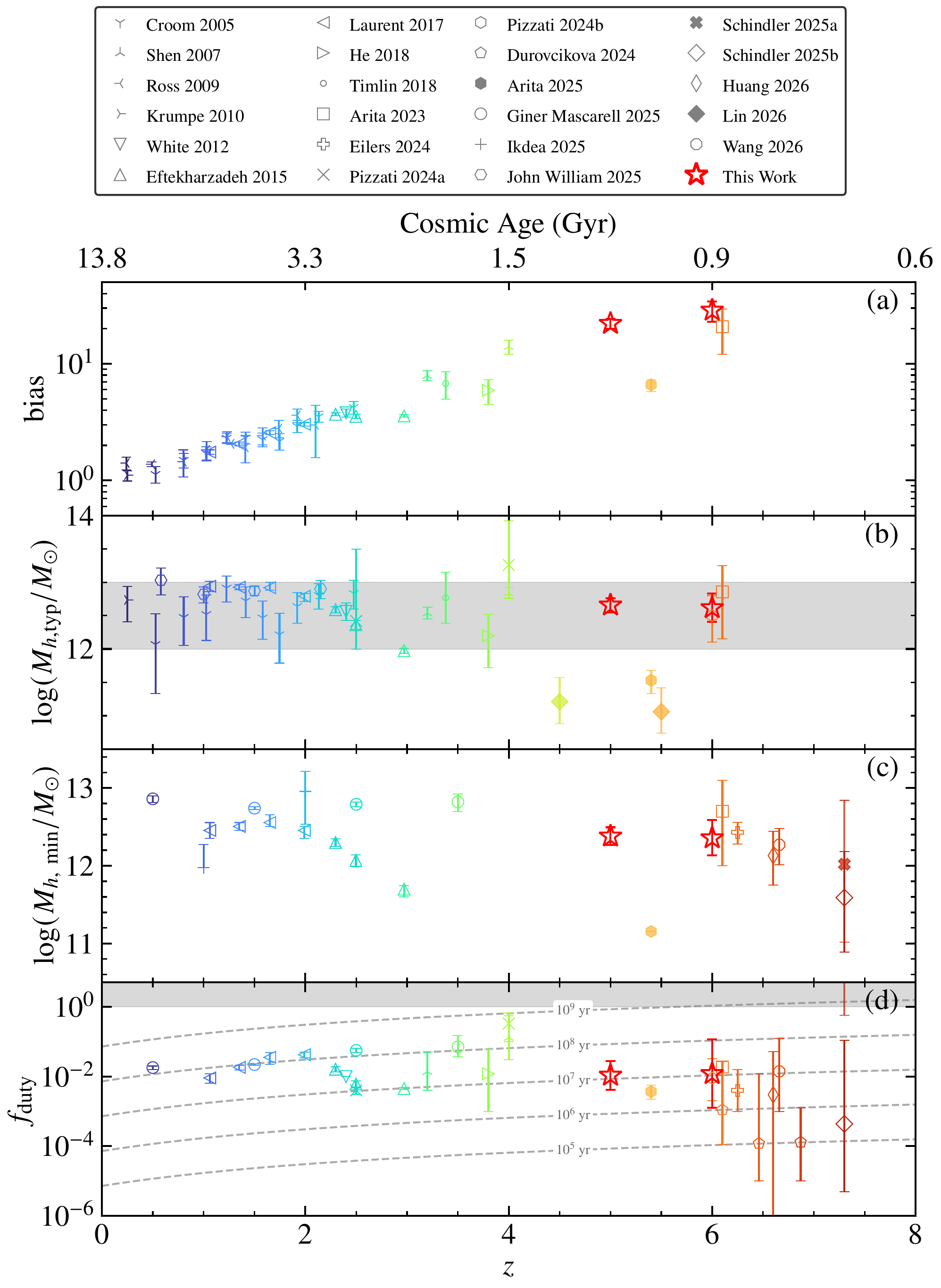}
    \caption{The  bias parameters (a), the typical DMH mass (b), the minimum DMH mass (c), and the duty cycle (d) based on clustering analyses from $z \sim 0$ to $z\sim 7.3$.  The  quasar results (hollow points) are extracted from \citet{DMHM_Croom_2005,DMHM_Shen_2007,Bias_Ross_2009,DMHM_Krumpe_2010,DMHM_White_2012,DMHM_Eftekharzadeh+2015,DMHM_Laurent_2017,DMHM_He_2018,DMHM_Timlin_2018,DMHM_Arita_2023,duty_cycle_Ďurovčíková_2024,DMHM_Eilers_2024,Pizzati2024a, Pizzati2024b, DMHM_Ikeda_2025,DMHM_Giner_Mascarell_2025,JT2025,DMHM_John_William_2025,Huang_2026_DMHM,Wang_2026_DMHM} and this work. The AGN and LRD results (filled points) are extracted from \citet{DMHM_Arita_2025,DMHM_Lin_2026}, and \citet{DMHM_Schindler_2025}, respectively. Grey shaded areas denote: (b) the typical DMH mass range of quasars; (d) the physically forbidden region for $f_{\rm duty}$.} 
    \label{fig:comparison}
\end{figure*}

It is challenging to investigate the potential evolution of the DMH mass of quasars across cosmic time. On the one hand, it is difficult to compare analogous quasar populations across cosmic time $-$ whether by matching quasars with the same luminosity (while controlling for other properties) at different epochs, or by comparing quasars with the same luminosity ranking within the full quasar sample at each respective epoch. As will be discussed below, even matching quasars with the same luminosity would lead to different estimates. On the other hand, even at the same cosmic epoch, quasar samples from different observational studies may differ significantly in their intrinsic properties, further complicating any straightforward evolutionary interpretation.

We compare our estimates of the bias parameters, typical DMH masses, minimum DMH masses as well as the duty cycle with values from the literature across different cosmic epochs, in Figure \ref{fig:comparison}. Caution is warranted, as the methods used to derive these quasar properties vary across studies, and the adopted cosmological models may also differ. Both factors can lead to slight discrepancies in the reported bias parameters, DMH masses, and duty cycles. Furthermore, some studies provide only the typical or minimum halo mass of quasars. For those that derive DMH masses from the bias parameter, we do not recompute the masses based on the cosmological parameters adopted in this work; instead, we directly present their original DMH mass estimates.

At $z\sim5.0$, our estimated bias parameter ($22.12^{+2.19}_{-2.21}$) is significantly higher than that of low-luminosity AGNs ($6.61^{+0.71}_{-0.82}$) reported by \citet{DMHM_Arita_2025}, indicating that bright quasars are more strongly clustered and reside in denser large-scale environments than their low-luminosity counterparts. At $z\sim6$, our estimated bias parameter ($28.70^{+5.56}_{-5.74}$) is consistent with the value found by \citet{DMHM_Arita_2023}, providing further evidence for a clear increase in the bias parameter from $z=0$ to $z\sim6$.

The characteristic DMH mass of luminous quasars is $\sim10^{12-13}M_{\odot}$, remaining almost constant with redshift up to $z\sim6$ within the uncertainties. For instance, \citet{DMHM_Arita_2023} measure $M_{h,\rm typ} = 5.0_{-4.0}^{+7.4} \times 10^{12} h^{-1}M_{\odot}$ for quasars with a much lower typical luminosity ($M_{1450}\sim-24.5$), which is similar to our estimates for more luminous quasars with median $M_{1450}\sim-25.9$ at $5.6 \leq z <6.3$. Using a large-volume $N$-body simulation, \citet{Pizzati2024b} obtain $\log_{10}(M_{h,\rm typ}/M_{\odot}) = 12.53\pm0.13$ for luminous quasars at $z\approx6.0$ with bolometric luminosities comparable to those in \citet{DMHM_Eilers_2024}. These results collectively suggest that the typical halo mass of luminous quasars remains roughly constant across a wide range of redshifts, despite the different methodologies and samples involved.

\subsection{Luminosity and Halo Mass Scaling Relation} 


\begin{figure*}[!ht]
    \centering
    \includegraphics[width=0.9\linewidth]{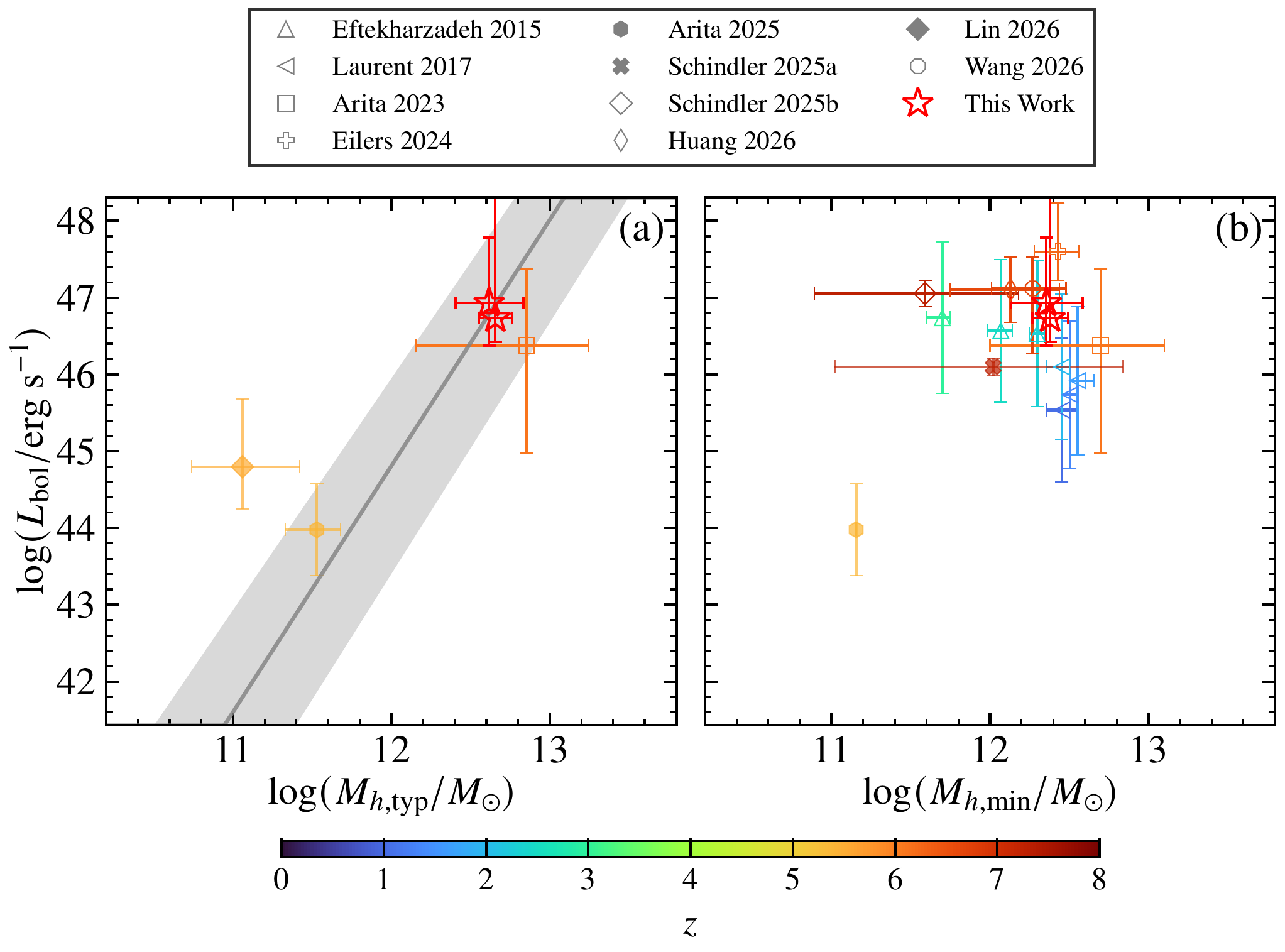}
    \caption{The scaling relation between the observed quasar bolometric luminosity and its derived host halo mass (a: typical mass; b: minimum mass). The errorbars of $L_{\rm bol}$ represent the observed range of the quasar luminosity, while the errorbars of $M_{h}$ denote the measurement uncertainties. Gray solid line and shaded region in panel a represent the theoretical $L_{\rm bol}-M_{\rm typ}$ relation at $z\approx6$ from simulations \citep{Pizzati2024b}.}
    \label{fig:Lbol-M_typ Relation}
\end{figure*}

DMH mass is a fundamental parameter governing the growth of galaxies and SMBHs. It determines the depth of the gravitational potential well, thereby influencing gas accretion, galaxy assembly, and the large-scale clustering properties of the quasar/AGN population. However, the observed luminosity of a quasar or AGN is not determined solely by the mass of its host halo. Additional factors, including the black hole mass, Eddington ratio, radiative efficiency, and the duty cycle of nuclear activity, can introduce substantial scatter in the connection between luminosity and halo mass. To investigate whether the most luminous quasars preferentially reside in the most massive DMHs, we compile measurements from the literature and compare the bolometric luminosities of quasars and AGNs with their inferred characteristic halo masses and minimum halo masses.

For quasars or AGNs, different studies evaluate the intrinsic luminosity using different tracers, including the H$\alpha$ luminosity ($L_{\rm H\alpha}$) or rest frame $M_{1450}$. To enable a consistent comparison, we uniformly convert these measurements into bolometric luminosity, $L_{\rm bol}$. For sources with available $L_{\rm H\alpha}$, we derive $L_{\rm bol}$ using the scaling relation from \citet{Lbol_LHalpha_Greene_2007}:
\begin{equation}
L_{\rm bol} = 2.34\times10^{44} \left(\frac{L_{\rm H\alpha}}{10^{42}}\right)^{0.86}~\rm erg~s^{-1}.
\end{equation}

For studies that present $M_{1450}$, we convert them to $L_{\rm bol}$ according to \citet{M1450_to_Lbol_Runnoe_2012}:

\begin{align}
    M_{1450} &= -2.5 \log_{10}\!\left[f_\nu(10\,\mathrm{pc})\right] - 48.60, \\
    L_\nu(1450\,\text{\AA}) &= 4\pi (10\,\mathrm{pc})^2 \cdot  f_\nu(10\,\mathrm{pc}), \\
    L_{\mathrm{bol}} &= 4.2 \cdot  \nu \, L_\nu(1450\,\text{\AA}),
\end{align}

As shown in Figure \ref{fig:Lbol-M_typ Relation}, the characteristic (minimum) DMH mass required to host a luminous quasar with a typical bolometric luminosity of $10^{46-47}~\mathrm{erg\,s^{-1}}$ at $z\sim6$ is remarkably consistent across different studies, clustering around $\sim10^{12.5}~M_\odot$ ($\sim10^{12.2}~M_\odot$), despite the substantial uncertainties associated with individual measurements. In contrast, lower-luminosity quasars and AGNs with typical bolometric luminosities of $10^{44-45}~\mathrm{erg\,s^{-1}}$ are generally hosted by significantly less massive halos, with characteristic (minimum) halo masses of approximately $10^{11.2}~M_\odot$ ($10^{11.0}~M_\odot$). Overall, quasar bolometric luminosity exhibits a strong positive correlation with host halo mass over nearly three orders of magnitude in luminosity, although the scatter remains considerable at fixed luminosity. This trend suggests that more luminous quasars preferentially reside in more massive DMHs, whose deeper gravitational potential wells can support larger gas reservoirs and sustain the high accretion rates required to power the observed quasar luminosities. The substantial scatter further indicates that halo mass is not the sole factor governing quasar luminosity; variations in black hole mass, Eddington ratio, radiative efficiency, and duty cycle likely also play important roles.

The $L_{\rm bol}$--$M_h$ scaling relation at $z\approx6$ has been investigated by \citet{Pizzati2024b} using the FLAMINGO cosmological simulations \citep{FLAMINGO_2023}. Their model combines the quasar luminosity function from \citet{JT2023}, clustering constraints from \citet{DMHM_Arita_2023}, and a simple power-law relation between quasar bolometric luminosity and DMH mass. Despite its simplicity, the model successfully reproduces the available observational constraints over nearly four orders of magnitude in luminosity, spanning from relatively faint AGNs ($L_{\rm bol}\sim10^{43.5}~\mathrm{erg\,s^{-1}}$) to the most luminous quasars ($L_{\rm bol}\sim10^{47.5}~\mathrm{erg\,s^{-1}}$). Our measurements are broadly consistent with this relation within the current uncertainties and further extend the observational constraints at the luminous end, as presented in Figure \ref{fig:Lbol-M_typ Relation}. Future clustering measurements covering a broader range of luminosities and redshifts, particularly at the highest redshifts where current constraints remain sparse, will be crucial for refining the connection between black hole growth and DMH assembly in the early Universe.


\subsection{Comparison of Quasar Duty Cycles}

\begin{figure*}[!ht]
    \centering
    \includegraphics[width=0.9\linewidth]{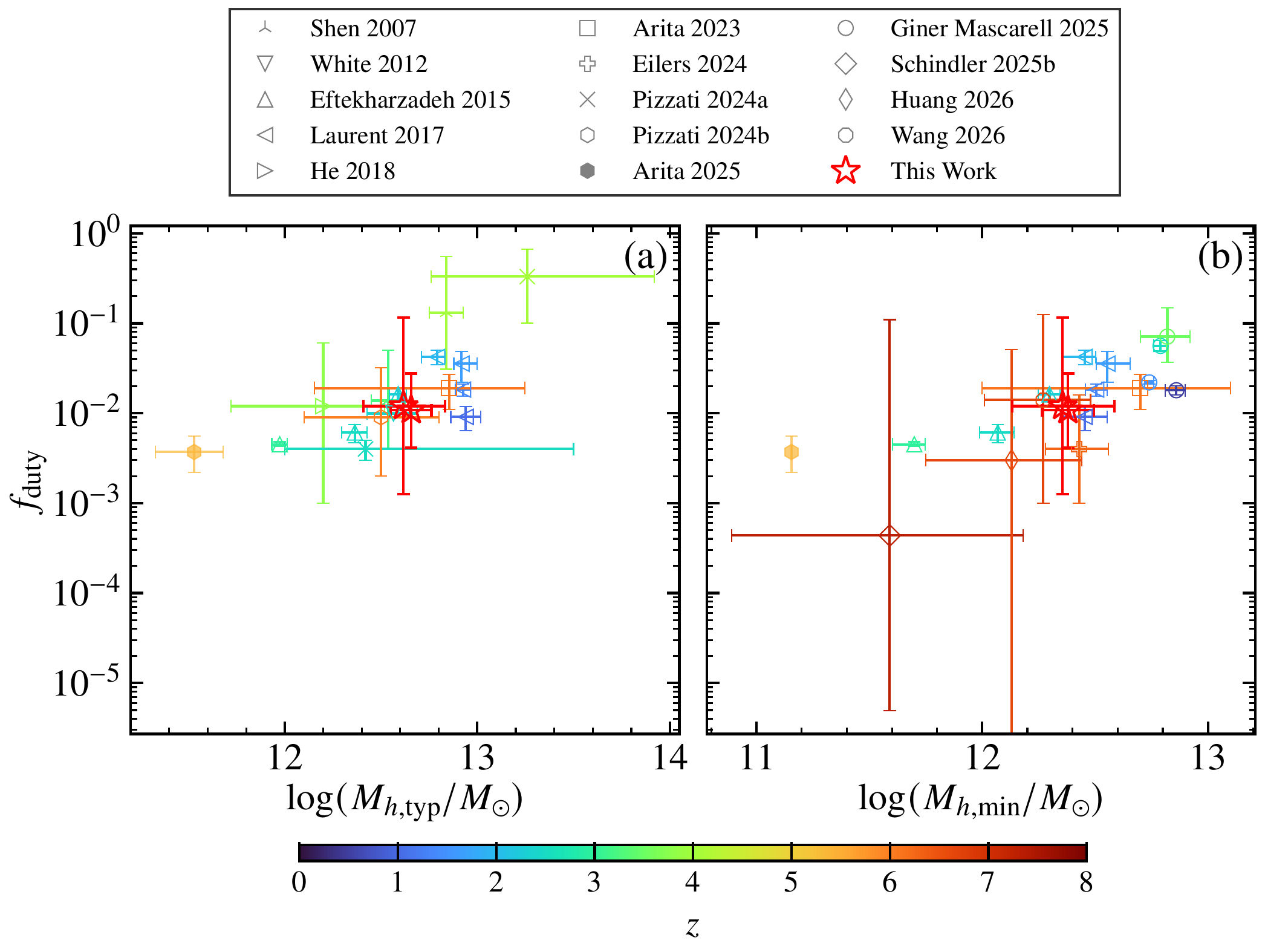}
    \caption{The relationship between the quasar duty cycle and its host halo mass across a wide redshift range. Panel (a) shows the duty cycle as a function of the typical halo mass, while panel (b) shows it as a function of the minimum halo mass.}
    \label{fig:f_duty-M_typ Relation}
\end{figure*}

As shown in Figure \ref{fig:comparison}, the inferred quasar lifetimes span a wide range across cosmic time. Most measurements at $z\lesssim4$ suggest characteristic lifetimes of $t_{\rm Q}\sim10$--$100~\mathrm{Myr}$, although substantial scatter exists among different studies due to variations in sample selection, analysis techniques, and underlying assumptions. In contrast, observational constraints in the redshift interval $4\lesssim z\lesssim6$ remain relatively limited, with only a handful of measurements currently available, including those of \citet{DMHM_Arita_2025} at $z\sim5.5$ and our measurement at $z\sim5.0$. At $z\approx6$, our inferred duty cycle and corresponding quasar lifetime are broadly consistent with previous clustering-based estimates \citep{DMHM_Arita_2023,DMHM_Eilers_2024,Pizzati2024b,duty_cycle_Ďurovčíková_2024}. Together, these studies indicate characteristic quasar lifetimes of order $10$--$30~\mathrm{Myr}$ at the ending phase of the reionization epoch. At even higher redshifts ($z\gtrsim6.5$), both duty-cycle and lifetime estimates become considerably more uncertain, primarily owing to the limited number of known quasars and the consequently weak constraints on their clustering properties. Future wide-area surveys are expected to substantially improve these measurements by increasing the available sample sizes and reducing the statistical uncertainties.

We present the relationship between quasar duty cycle and host halo mass in Figure \ref{fig:f_duty-M_typ Relation}. Despite the substantial scatter among individual measurements, a clear positive correlation is evident over the wide redshift range $0\lesssim z\lesssim8$, with quasars hosted by more massive DMHs generally exhibiting higher duty cycles. Our measurements at $z\sim5$ and $z\sim6$, corresponding to duty cycles of $1.1\%$ and $1.2\%$, respectively, follow the same overall trend and are fully consistent with previous observational constraints. These duty cycles imply characteristic quasar lifetimes of several tens of Myr, sufficiently long to sustain significant SMBH growth during a single active episode. The observed correlation likely reflects the fact that more massive halos are able to provide larger gas reservoirs and maintain the conditions required for prolonged or recurrent accretion onto the central black hole. The consistency between our measurements and the global $f_{\rm duty}$--$M_h$ relation further suggests that the connection between quasar activity and DMH mass was already in place within the first billion years after the Big Bang.

The observed correlation can be naturally understood within the framework of halo-regulated black hole growth. More massive DMHs possess deeper gravitational potential wells and larger reservoirs of cold gas, enabling more sustained and potentially more frequent accretion episodes onto the central SMBH. Moreover, galaxy interactions and mergers are expected to occur more frequently in massive halos, promoting gas inflows toward the nuclear region and increasing the likelihood of triggering luminous quasar activity. As a result, the probability that a halo hosts an active quasar at any given epoch is expected to increase with halo mass.

A notable feature of Figure~\ref{fig:f_duty-M_typ Relation} is that measurements spanning nearly the entire observable history of the Universe occupy a relatively narrow locus in the $f_{\rm duty}$--$M_h$ plane. Although some redshift evolution is apparent, the dependence of duty cycle on halo mass appears to be stronger than its dependence on cosmic epoch. This behavior suggests that host halo mass is a primary parameter governing the occurrence of luminous quasar activity, while cosmic evolution mainly modulates the normalization of the relation through changes in gas supply, merger rates, and the efficiency of black hole fueling. The existence of a relatively tight $f_{\rm duty}$--$M_h$ relation over a broad redshift range further implies that the fundamental connection between DMH assembly and black hole growth has remained largely in place since the early stages of galaxy formation.

\section{Conclusions}\label{sec:Conclusions}

Using a sample of 1,242 spectroscopically confirmed high-redshift quasars and 827 highly reliable photometric quasar candidates at $4.7 \leq z < 6.3$ (excluding $5.4 < z < 5.6$), we perform clustering analyses of high-redshift quasars in this redshift range. The median $M_{1450}$ of these quasars is $\sim -25.4$ for $4.7 \leq z <5.4$ and $\sim -25.9$ for $5.6 \leq z <6.3$. From the clustering analyses, we derive a variety of quasar properties, including their DMH masses, bias parameters and duty cycles. Our main conclusions are summarized as follows:
\begin{enumerate}
    \item We measure the projected auto correlation function of high-redshift quasars and find that it is well described by a real-space power-law correlation function of the form $\xi(r)=(r/r_0)^{-\gamma}$, indicating strong large-scale clustering. Fitting the projected auto-correlation function yields a correlation length of $r_0=35.49^{+4.29}_{-4.59}\ h^{-1}\,\mathrm{Mpc}$ with a fixed slope of $\gamma=1.6$ for the redshift interval $4.7 \leq z < 5.4$, and $r_0=42.85^{+9.63}_{-11.32}\ h^{-1}\,\mathrm{Mpc}$ with $\gamma=1.6$ for $5.6 \leq z < 6.3$.
    
    \item We estimate the bias parameter and the typical DMH mass. The DMH mass is evaluated as $\log{(M_h/M_\odot}) = 12.66^{+0.11}_{-0.11}$ with the bias parameter, $b=22.12^{+2.19}_{-2.21}$ for quasars at $4.7 \leq z <5.4$. And the DMH mass is estimated to be $\log{(M_h/M_\odot}) = 12.62^{+0.21}_{-0.21}$ with the bias parameter, $b = 28.70^{+5.56}_{-5.74}$ for quasars at $5.6 \leq z <6.3$. 

    \item We estimate the quasar duty cycle $f_{\rm duty}$ for both redshift intervals of $4.7 \leq z <5.4$ and $5.6 \leq z <6.3$ by adopting the current QLF measurements and a DMH mass function, obtaining $\log(M_{h,\min}/M_{\odot})=12.38^{+0.11}_{-0.11}$  and $f_{\rm duty} = 0.011^{+0.017}_{-0.007}$ for quasars at $4.7 \leq z <5.4$, and $\log(M_{h,\min}/M_{\odot})=12.36^{+0.23}_{-0.22}$  and $f_{\rm duty} = 0.012^{+0.104}_{-0.011}$ for quasars at $5.6 \leq z <6.3$. These comparably small duty cycle estimates might indicate that a significant fraction of SMBH growth occurs in an obscured phase. 

    \item Independent clustering analyses based on the spectroscopic-only and combined quasar samples yield statistically consistent results for all derived quantities, including the $r_0$, $b$, $M_{\rm typ}$, $M_{\rm min}$, $f_{\rm duty}$, and $t_{\rm Q}$. The inclusion of high-confidence photometric candidates substantially improves the statistical precision of the clustering measurements while preserving the inferred physical properties of the quasar population within the uncertainties. This close agreement demonstrates that the photometric candidates are highly reliable and provides strong validation of the candidate-selection procedure adopted in this work.
    
    \item Although comparisons among different clustering measurements are inevitably subject to systematic uncertainties arising from sample selection, analysis techniques, and halo-mass modeling, the overall evolutionary trend provides valuable insight into the connection between quasars and their host DMHs across cosmic time. The bias parameter exhibits a pronounced increase from $z\sim0$ to $z\sim6$, reflecting the fact that luminous quasars at higher redshifts occupy increasingly rare density peaks in the underlying matter distribution. In contrast, the characteristic halo mass of luminous quasars remains remarkably constant, at $\log(M_{\rm typ}/M_\odot)\sim12$--13, over the same redshift range. This behavior suggests that the halo mass scale most conducive to efficient black hole growth and luminous quasar activity has changed little throughout cosmic history, even though halos of a given mass become progressively more biased at earlier epochs.

    \item We find that quasar bolometric luminosity is strongly correlated with host halo mass across a wide luminosity range, although significant scatter exists at fixed luminosity. The quasar duty cycle also increases systematically with halo mass and exhibits only weak redshift dependence. These results suggest that DMH mass is a primary factor governing quasar activity, with more massive halos providing the gas supply and environmental conditions necessary to sustain prolonged black hole growth and luminous quasar episodes.

\end{enumerate}

\section{Acknowledgments}

HM, HZ and GY acknowledge financial support from the start-up funding of the Huazhong University of Science and Technology, the National Science Foundation of China grant (No. 12303007)  and the China Manned Space Program (CMS-CSST-2025-A06). We thank the anonymous referees for their constructive suggestions, which  helps to improve our manuscript significantly. HZ designed the project,  offered help for data analysis and led results interpretation. GY led the data sample preparation and HM led the correlation analyses and halo properties deviation. HM and GY contribute equally. 

This research used data obtained with the Dark Energy Spectroscopic Instrument (DESI). DESI construction and operations are managed by the Lawrence Berkeley National Laboratory. This material is based upon work supported by the U.S. Department of Energy, Office of Science, Office of High-Energy Physics, under Contract No. DE–AC02–05CH11231, and by the National Energy Research Scientific Computing Center, a DOE Office of Science User Facility under the same contract. Additional support for DESI was provided by the U.S. National Science Foundation (NSF), Division of Astronomical Sciences under Contract No. AST-0950945 to the NSF’s National Optical-Infrared Astronomy Research Laboratory; the Science and Technology Facilities Council of the United Kingdom; the Gordon and Betty Moore Foundation; the Heising-Simons Foundation; the French Alternative Energies and Atomic Energy Commission (CEA); the National Council of Science and Technology of Mexico (CONACYT); the Ministry of Science and Innovation of Spain (MICINN), and by the DESI Member Institutions: www.desi.lbl.gov/collaborating-institutions. The DESI collaboration is honored to be permitted to conduct scientific research on Iolkam Du’ag (Kitt Peak), a mountain with particular significance to the Tohono O’odham Nation. Any opinions, findings, and conclusions or recommendations expressed in this material are those of the author(s) and do not necessarily reflect the views of the U.S. National Science Foundation, the U.S. Department of Energy, or any of the listed funding agencies.

Based on observations collected at the European Southern Observatory under ESO programme(s) 109.238W.003, 105.208F.001, and 0104.A-0812(A), and/or data obtained from the ESO Science Archive Facility with DOI(s) under https://doi.org/10.18727/archive/41.

%
\onecolumngrid               

\bibliographystyle{aasjournal}   
\bibliography{bibliography}

\end{document}